\documentclass[%
reprint,
amsmath,amssymb,
aps,
]{revtex4-2}

\usepackage{graphicx}
\usepackage{dcolumn}
\usepackage{bm}
\usepackage{hyperref}
\usepackage{url}
\hypersetup{
    colorlinks=true,
    linkcolor=blue,
    filecolor=magenta,      
    urlcolor=cyan,
    }

\begin{document}

\preprint{APS/123-QED}

\title{Atomistic Nature of Amorphous Graphite}

\author{C. Ugwumadu}
\email{cu884120@ohio.edu}
\affiliation{Department of Physics and Astronomy, \\
Nanoscale and Quantum Phenomena Institute (NQPI),\\
Ohio University, Athens, Ohio 45701, USA}%

\author{K. Nepal}
\affiliation{Department of Physics and Astronomy, \\
Nanoscale and Quantum Phenomena Institute (NQPI),\\
Ohio University, Athens, Ohio 45701, USA}%

\author{R. Thapa}
\affiliation{Department of Physics and Astronomy, \\
Nanoscale and Quantum Phenomena Institute (NQPI),\\
Ohio University, Athens, Ohio 45701, USA}%

\author{D. A. Drabold}%
\email{drabold@ohio.edu}
\affiliation{Department of Physics and Astronomy, \\
Nanoscale and Quantum Phenomena Institute (NQPI),\\
Ohio University, Athens, Ohio 45701, USA}%

\date{\today}

\begin{abstract}
This paper focuses on the structural, electronic, and vibrational features of amorphous graphite [R. Thapa \textit{et. al.}, Phys. Rev. Lett. 128, 236402 (2022)]. The structure order in amorphous graphite is discussed and compared with graphite and amorphous carbon. The electronic density of states and localization in these phases were analyzed. Spatial projection of charge densities in the $\pi$ bands showed a high charge concentration on participating atoms in  connecting hexagons. A vibrational density of states was computed and is potentially an experimentally testable fingerprint of the material. An analysis of the vibrational modes was carried out using the  phase quotient, and the mode stretching character. The average thermal conductivity calculated for aG was 0.85 Wcm$^{-1}$K$^{-1}$ and 0.96 Wcm$^{-1}$K$^{-1}$ at room temperature and 1000 K respectively.

\end{abstract}


\maketitle


\section{\label{sec:introduction}INTRODUCTION}
The growing and unmet industrial demand for graphite, coupled with the associated environmental problems resulting from graphite mining activities have become a critical issue \cite{flakesbook}. While the graphite feed-stock used for only lithium batteries was projected to reach an annual demand of 1.25 million tonnes by 2025 \cite{flakes1}, the total amount of mined graphite was only 1 million tonnes in 2021 \cite{USGS}. A promising method intended to mitigate the graphite supply shortage involves a "second-life" approach of graphite recycling/reuse from the spent lithium-ion batteries \cite{R1, R2, R3, R4, R5}. However, recent reports suggest that the environmental and economic implications of industrial-scale second-life graphite are still not favorable \cite{SL0, SL1, SL2, SL3}. Another area of tremendous research interest is the graphitization of naturally occurring carbonaceous materials like coal \cite{gp1,gp2,gp3,gp4,gp5,gp6}. Beyond the obvious ecological and economic benefits, the actualization of this form of modern-day alchemy would revolutionize the frontiers of science and engineering. Unfortunately, large-scale graphitization has not yet been achieved and any attempt to realize this would undoubtedly require a synergy between experiments and simulations. 

It has been suspected from experiments that graphitization occurs near 3000 K, but until recently, the details of the formation process and nature of the disorder in the planes remained unknown. Our recent prediction of amorphous graphite (aG) from \textit{ab initio} and machine learning molecular dynamic simulations suggested the possibility that the  material exists \cite{LAG}. We showed that carbon has an overwhelming tendency to layer, even with topological defects like 5- and 7- member rings, which fit quite naturally into the network. This discovery has fostered a renewed experimental interest in the path to synthetic forms of graphite from non-crystalline carbon structures. However, a detailed study of the atomistic nature of this carbon structure is required for significant new advances.

In this paper, we elucidate the structural, electronic, and vibrational properties of aG using an ensemble of model sizes ranging from 160 - 3200 atoms. We investigated the effects of the periodic boundary conditions (PBC) in the formation process of aG and compared the atomic structure of aG to graphite and amorphous carbon. We explored the electronic structure and vibrations by computing the density of states and their corresponding inverse participation ratio. Additionally for the phonon vibrations,  the phase quotient, and bond-stretching character were computed. We note here that, except stated otherwise, the analysis herein for the aG was compared with a pristine graphite model (pG) and low-density amorphous Carbon (aC) taken from references \cite{pG_EXAFS} and \cite{DAD_aC} respectively. Molecular dynamics calculations were done using the ``Vienna \textit{Ab initio} Simulation Package" (\texttt{VASP}) with plane-wave potentials \cite{VASP}, and the ``Large-scale Atomic/Molecular Massively Parallel Simulator" (\texttt{LAMMPS}) \cite{lammps} using the Machine-learning Gaussian Approximation Potential (ML-GAP) \cite{C}. Finally, a subscript ``n" which represents the number of atoms in the system will be used to define the amorphous graphite models (aG$_n$). 

\section{\label{sec:structure}Formation and Structure}
 A detailed description of the simulation protocol for aG can be found in ref. \cite{LAG}. In short, The aG formation process involves annealing of \textit{ab initio} models of amorphous carbon or a random starting configuration of carbon atoms within the ``formation density" range of ca. 2.2-2.8 g/cm$^3$ in a canonical (NVT) ensemble at temperatures ranging from 2700 $\sim$ 4000 K for up to 500 ps. For this work, we generated an ensemble of structural models (15 models for each aG$_n$) from different starting configurations at 3000 K and density of 2.44 g/cm$^3$. The temperature was controlled using the Nos\'e–Hoover thermostat as implemented within \texttt{VASP}  and \texttt{LAMMPS}. The animation for the aG formation process, provided in the supplementary material \cite{suppl}, indicates that aG is formed in a two-stage process. (1) Conversion of non-sp$^2$ into sp$^2$ coordination. (2) The separation of the layers of sp$^2$ atoms into amorphous graphene sheets. 

The formation of aG is also dependent on the periodic boundary condition (PBC) applied. We observed that for aG, the PBC must be applied in three dimensions. In another work, we reported on the formation of buckyonions from a random C network placed in a 3D vacuum, such that periodic boundary condition describes a system of isolated carbon clusters. In the same light, we found nanotubes by maintaining the PBC along the z-axis only (cylindrical symmetry) \cite{BO}.

\begin{figure}[!htpb]
	\centering
	\includegraphics[width=\linewidth]{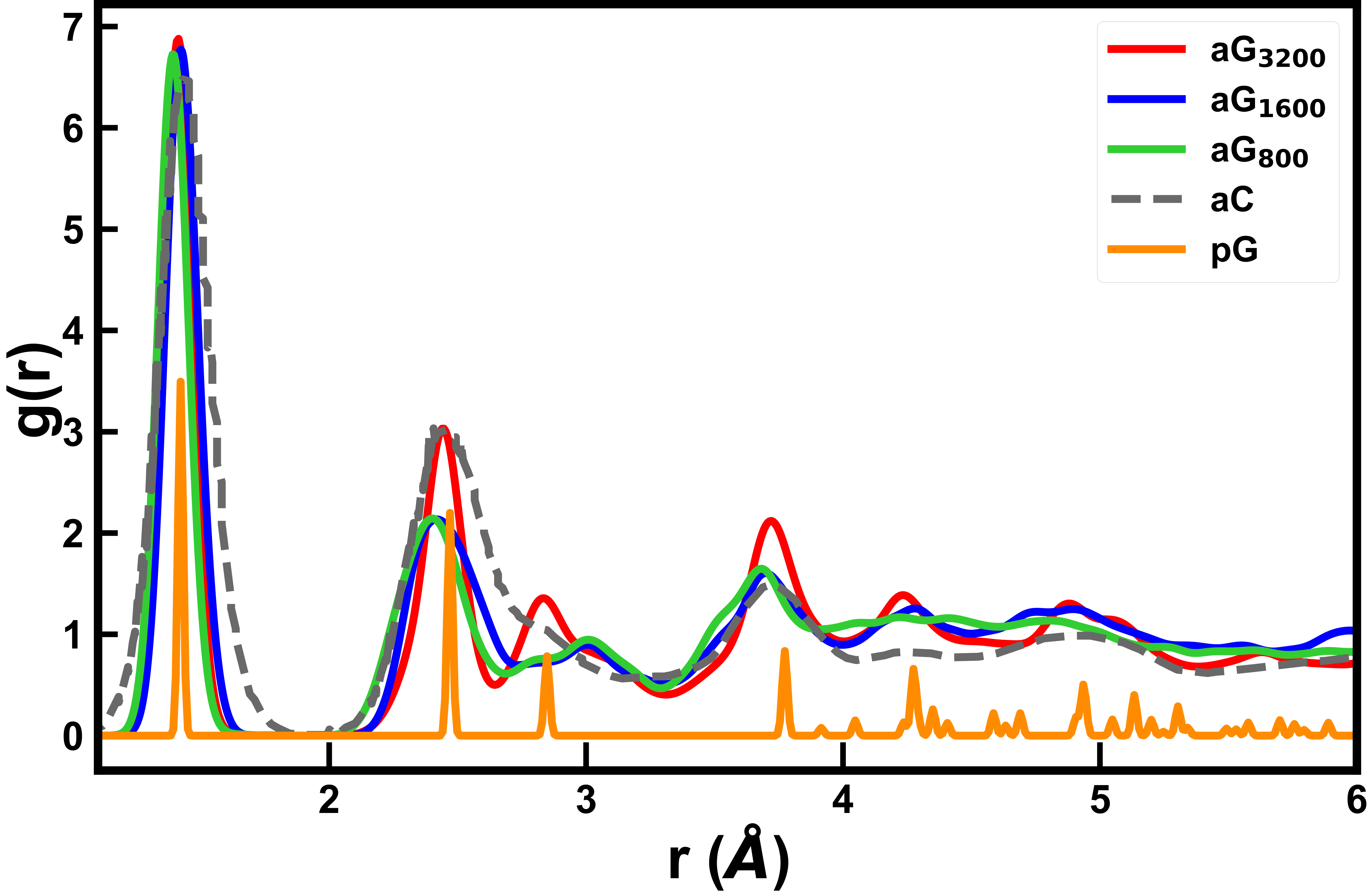}
    	\caption{Radial distribution function g(r) for different aG models compared with pG and  aC}
	\label{fig:Cfig_g(r)}
\end{figure}

The structural order of aG models was analyzed using pair correlation functions. Fig. \ref{fig:Cfig_g(r)} compares the peaks obtained for the aG models with pG and aC.  The first peak for all aG was within the nearest neighbor C-C bond length observed in graphite. The aG models reproduced more graphitic peaks as the system size increased. This is a consequence of the higher ratio of hexagonal to non-hexagonal rings (6:n; n = 5 or 7) found in large aG systems. 

\begin{figure}[!ht]
	\centering
	\includegraphics[width=\linewidth]{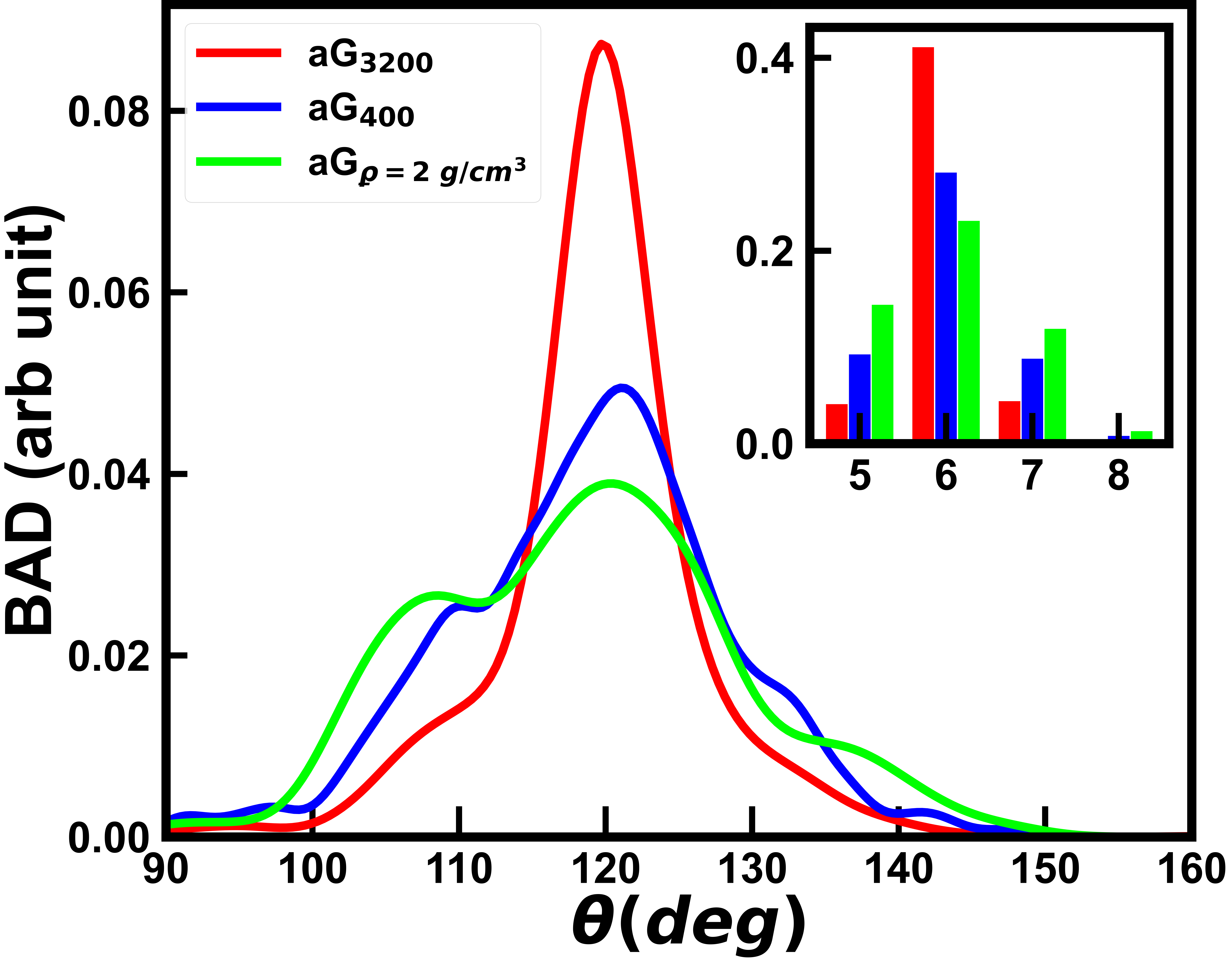}
    	\caption{Bond angle distribution (BAD) analysis and ring statistics (inset) for aG$_{3200}$ (red), aG$_{400}$ (blue) and a 160-atom aG model with a density of 2 g/cm$^3$ (green). The distribution has been smoothed to clearly show the peaks}
	\label{fig:Cfig_BAD}
\end{figure}

\begin{figure}[!ht]
	\centering
	\includegraphics[width=\linewidth]{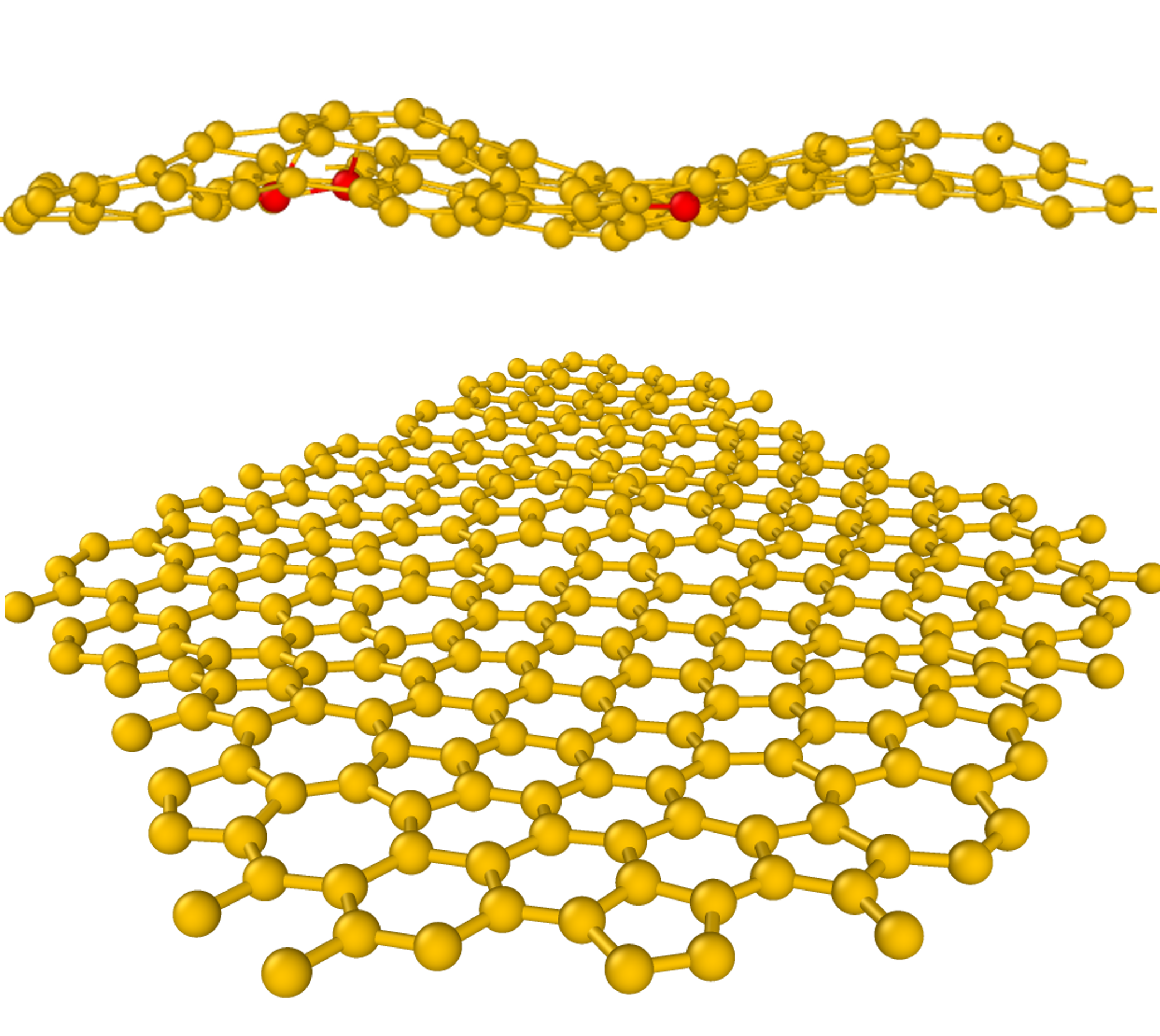}
    	\caption{Figure showing a representative layer for [TOP] the undulating (worm-like) structure ($\rho$ = 2 g/cm$^3$) and [BOTTOM] flat structure for aG formed below and within the desired density ($\rho$ = 2.44 g/cm$^3$) respectively. Yellow (red) represents three-fold (four-fold) coordination.}
	\label{fig:Cfig_aGplanes}
\end{figure}

 The 6:n ring ratio for the models was further confirmed from the bond angle distribution (BAD) and ring statistics. Fig. \ref{fig:Cfig_BAD} shows the C-C-C angle distribution in aG$_{3200}$ (red), aG$_{400}$ (blue), and aG formed at a lower ``threshold" density of 2.0 g/cm$^3$ (green). We will henceforth refer to the aG formed at 2.0 g/cm$^3$ as aG$_{<\rho}$. The BAD curve for aG$_{3200}$ (aG$_{400}$)  showed a sharp (flattened) peak at $\approx$ 120$^{\circ}$. The broad peak at $\approx$ 109$^{\circ}$ relates to non-hexagonal rings in the matrix. The ring statistics in the inset of Fig. \ref{fig:Cfig_BAD} confirmed that aG$_{3200}$ has a higher 6:n ring ratio when compared to the other two models. Importantly, Fig. \ref{fig:Cfig_BAD} (inset) indicates that the ratio of the 5- to 7- member rings (5:7) in aG$_{400}$ and aG$_{<\rho}$ is equal to and greater than unity respectively. In graphite-like structures with topological defects (i.e aG), planarity is achieved only if the positive curvature induced by a pentagonal ring is compensated by a negative curvature from a neighboring heptagon (or octagon) ring \cite{Mackay, TERRONES, Lenosky}. Deviation from a 1:1 ratio of pentagons and heptagons for an indeterminate number of hexagonal rings results in a complicated structure like the undulating, ``worm-like" layers seen in aG$_{<\rho}$ (see Fig. \ref{fig:Cfig_aGplanes} [TOP]), as opposed to  "almost" flat layers observed in aG within the desired density (see Fig. \ref{fig:Cfig_aGplanes} [BOTTOM]).

\begin{figure}[!htpb]
	\centering
	\includegraphics[width=\linewidth]{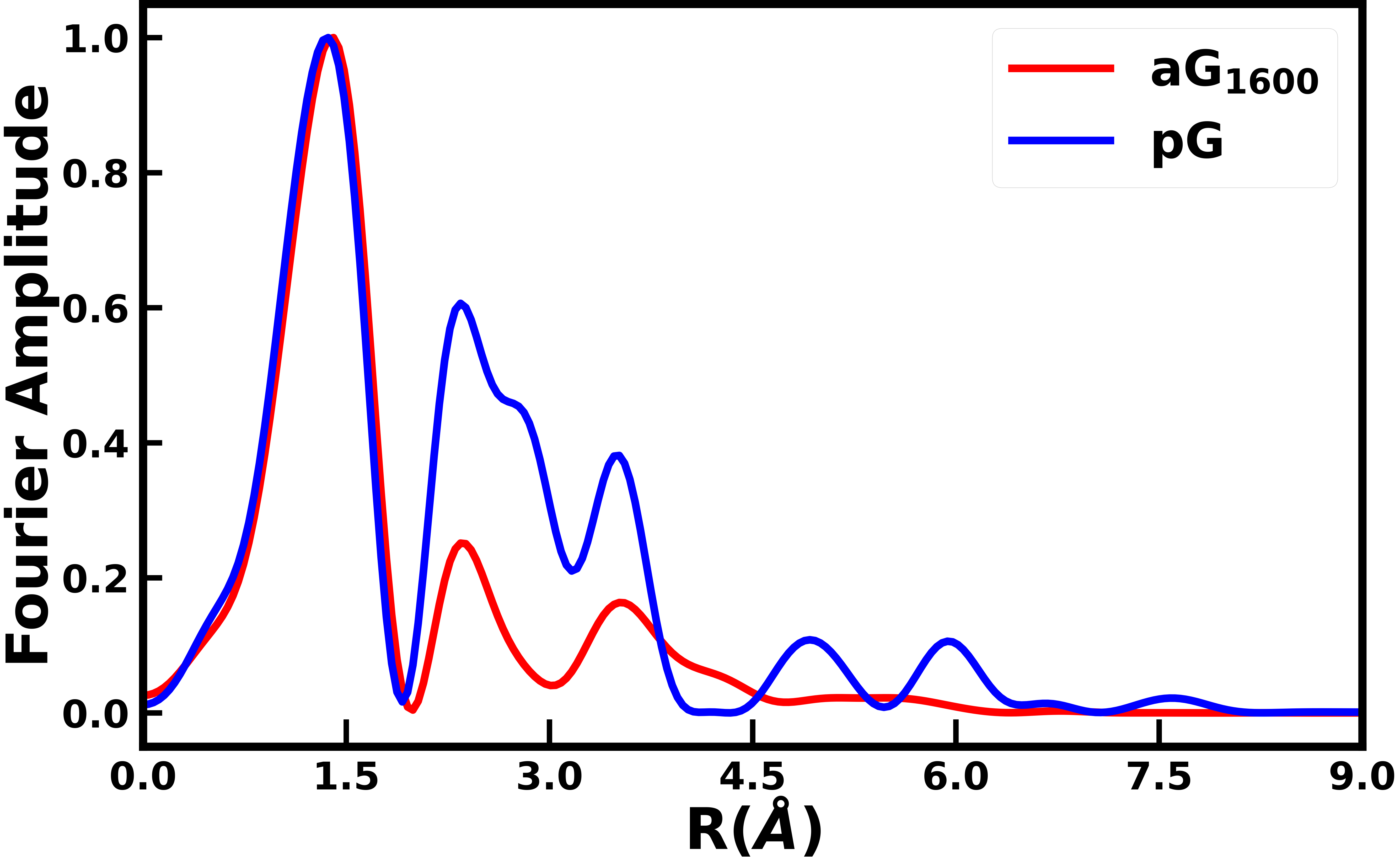}
    	\caption{Normalized Fourier amplitude for the C K-edge EXAFS spectra for pG and aG$_{1600}$.}
	\label{fig:Cfig_EAXFS_LAG}
\end{figure}

Next, we analyzed the local conformation and coordination number (N) of the aG models by implementing an \textit{ab initio} multiple scattering calculations of the extended x-ray absorption fine structure (EXAFS) using the real-space Green's function code FEFF10 \cite{FEFF10} for the K-edge. Using the Kaiser windowing function, with $\beta = 2$ \cite{kaiser}, the extracted post-edge oscillations ($\chi (E)$) were Fourier transformed (FT) into frequency space, and the resulting spectrum gives the radial distribution function (RDF). Fig. \ref{fig:Cfig_EAXFS_LAG} compares the normalized Fourier amplitude  acquired for aG$_{1600}$ and pG. All the peaks in aG$_{1600}$ corresponded with some peaks in pG. The first peak at 0.134 nm is due to the first-neighbor C-C scattering (0.142 nm, N = 3) \cite{comelli}. The second (0.243 nm, N = 6) and third (0.281 nm, N = 3) peaks in pG were resolved as a single second peak in aG at 0.243 nm. This "second peak" in aG has been identified in low-density amorphous carbon by Bhattarai and co-workers \cite{DAD_aC}. Unlike pG, the aG model did not produce additional peaks beyond the fourth C-C scattering peak at 0.38 nm, and this suggests an intermediate-range order in aG. We note here that the FT peaks calculated for pG are consistent with those published in literature \cite{comelli,Tanaka, Buades} and the results for aG are in agreement with the pair correlation function calculations in Fig. \ref{fig:Cfig_g(r)}. This presents a prediction to be employed with experiments.

\section{\label{sec:electronics} Electronic Structure}

The electronic density of states (EDoS) for aG was computed within \texttt{VASP} and the extent of localization of Kohn-Sham states ($\phi$) was calculated as the electronic inverse participation ratio (EIPR) using the following equation:

\begin{equation}
    I(\phi_n) =  \frac{\sum_i {| a_n^i |^4  }}{(\sum_i {| a_n^i |^2  } )^2}
    \label{eqn:EDoS}
    \end{equation}

\noindent where a$_n^i$ is the contribution to the eigenvector ( $\phi_n $) from the i$^{th}$ atomic orbital. High (low) values of EIPR indicate localized (extended) states. 

\begin{figure}[!htpb]
	\centering
	\includegraphics[width=\linewidth]{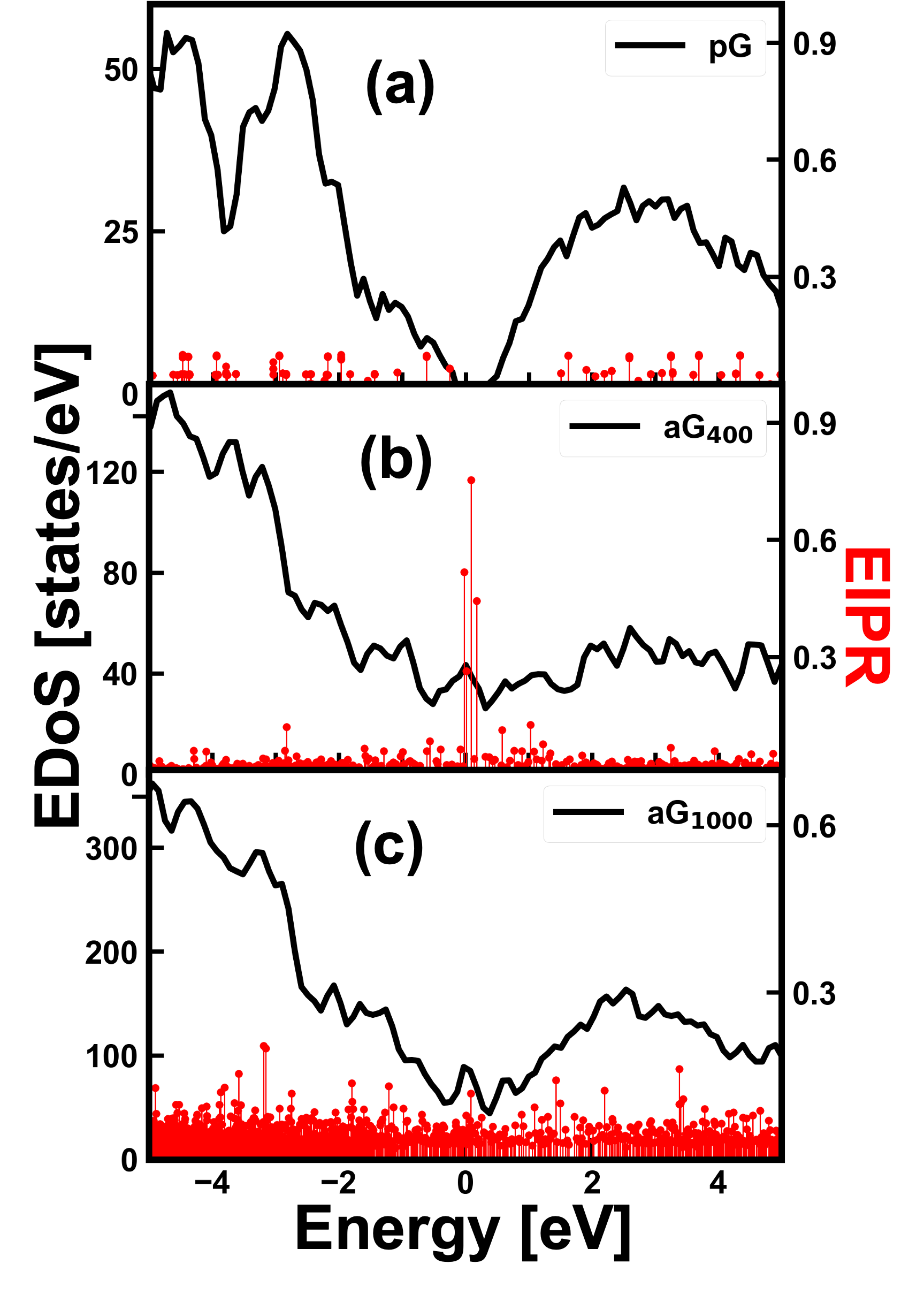}
    	\caption{Electronic DoS and IPR for pG and some aG models. }
	\label{fig:Kfig_edos}
\end{figure}

\begin{figure}[!htpb]
	\centering
	\includegraphics[width=.7\linewidth]{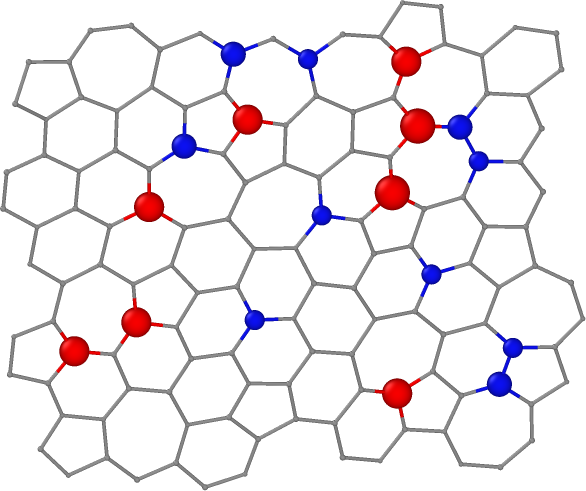}
    	\caption{Spatial projection of localized states near $E_f$ on the atoms in aG$_{400}$. Two states with the highest EIPR in Fig. \ref{fig:Kfig_edos} (b) are projected as Red- and blue-colored spheres}
	\label{fig:Kfig_localization}
\end{figure}

\noindent The EDoS and EIPR for pG, aG$_{400}$ and aG$_{1000}$ are shown in Fig. \ref{fig:Kfig_edos} (a-c) with the Fermi-level ($E_f$) shifted to zero. Fig. \ref{fig:Kfig_edos} (a) depicts an expected gap at $E_f$ in pG, with low EIPR values for states in the conduction and valence region \cite{DRABOLD1995833}. On the other hand, aG does not show any gap at $E_f$, and some states are localized (see \ref{fig:Kfig_edos} (b and c). The states with the highest EIPR values were predominantly distributed among non-hexagonal rings in the matrix as shown in Fig. \ref{fig:Kfig_localization}.

 
\begin{figure}[!htpb]
\includegraphics[width=\linewidth]{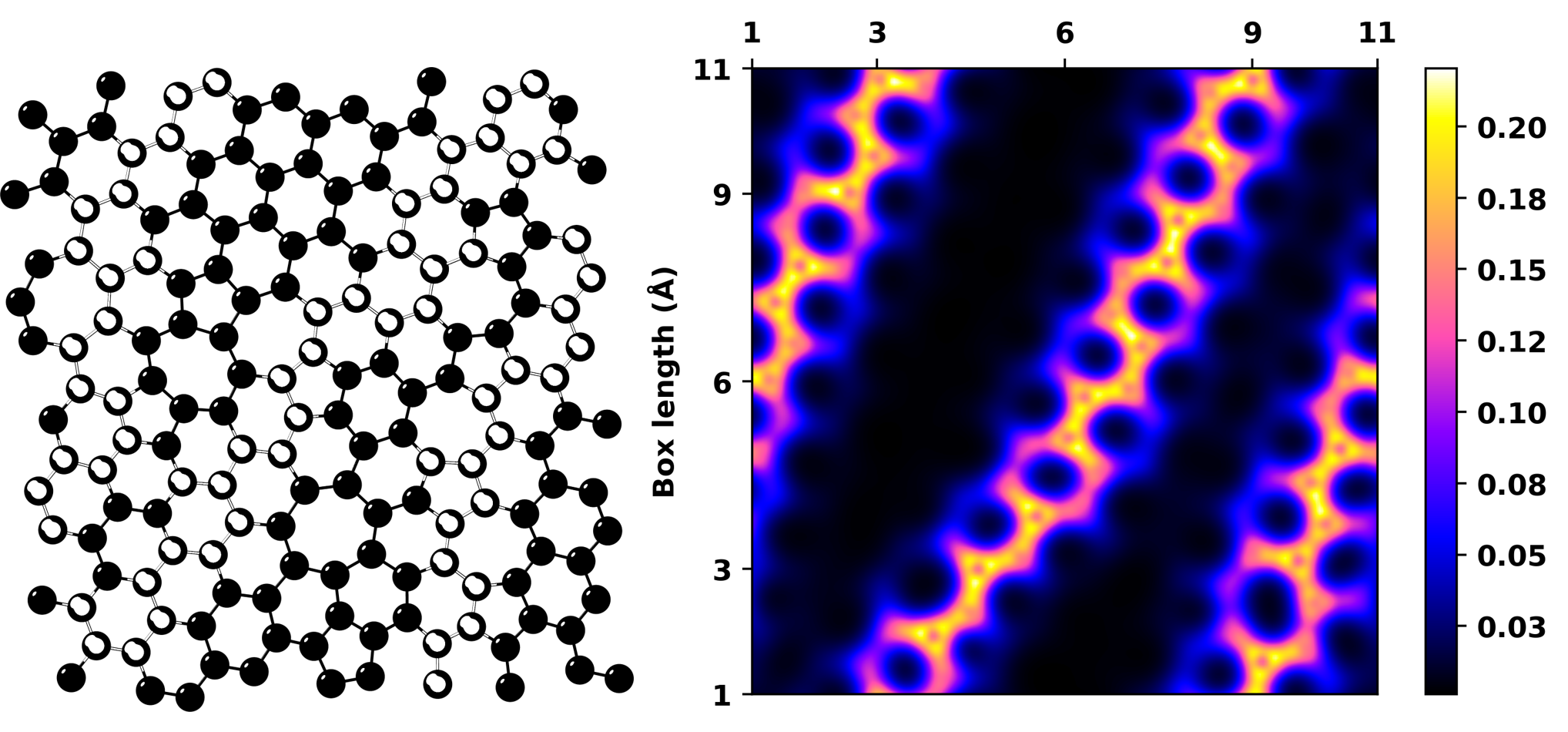}
\caption{ The $\pi$ orbital charge density distribution projected on a plane of aG. The white-colored atoms in the [LEFT] figure indicate the connected hexagonal path of the charge density [RIGHT]}
\label{fig:Cfig_CHGDensity}
\end{figure}

In ref. \cite{LAG}, using the space-projected conductivity (SPC) formalism \cite{SPC}, we showed that the conduction-active path in aG was exclusively along connecting atoms in hexagonal rings. To further develop this, we projected the laterally averaged charge density for the $\pi$ orbitals onto the planes of atoms. Our result, presented as a contour heat-map plot in Fig. \ref{fig:Cfig_CHGDensity} [RIGHT],  revealed that the regions with the highest values are on the planes with highly connected hexagonal rings (see illustration in Fig. \ref{fig:Cfig_CHGDensity} [LEFT]. This is consistent with our initial findings from the SPC calculation and also suggests that, even with the topological disorder in its layers, to some extent, aG possesses a degree of order in the way the electrons interact in-plane ($\sigma$ electrons) and out-of-plane ($\pi$ electrons).

\section{\label{sec:vibrations} Vibrations}

Thermal and mechanical properties, as well as the local bonding environment, can be obtained from the vibrations of amorphous materials. The harmonic approximation for vibrations requires evaluating the Hessian matrix, which is accomplished by force calculation from 0.015 \AA~atomic displacements in six directions ($\pm~x,\pm~y, \pm~z$). The vibrational density of states (VDoS) is calculated as:

\begin{equation}
    g(\omega) = \frac{1}{3N} \sum_{i=1}^{3N} \delta(\omega - \omega_i) 
\end{equation}

\noindent where, $N$ and $\omega_i$ represent the number of atoms and the eigen-frequencies of normal modes, respectively. The delta function (approximated by a Gaussian with a standard deviation equal to 1.5\% the maximum frequency) ensures that high-density values were assigned to vibration frequencies that lie close to the normal modes. The extent of localization of each normal mode frequency was calculated through the vibration inverse participation ratio (VIPR), defined as:

\begin{equation}
    V(\omega_n) = \frac{\sum_{i=1}^{N} |\boldsymbol{u}^{i}_{n}|^4}{(\sum_{i=1}^{N} |\boldsymbol{u}^{i}_{n}|^2)^2}  
    \label{eq:VIPR}
\end{equation}

\noindent where, $\boldsymbol{u^{i}_{n}}$ is displacement vector of i$^{th}$ atom at normal mode frequency $\omega_n$. By definition, low values of VIPR indicate vibrational mode evenly distributed among the atoms while higher values imply that few atoms contribute at that particular eigen-frequency.

Fig. \ref{fig:Cfig_VDOS_VIPR} shows the total VDoS for amorphous and pristine graphite [TOP] and  the extent of localization from the VIPR [BOTTOM]. While the peaks for aG and pG do not match, the overall shape of both models remained consistent. The figure also provides a vibrational fingerprint of aG to be examined in experiments. The VIPR for aG suggests that more states are localized in the high-frequency region which corresponds to the ``optical" mode. The classification of phonon vibrations into pure acoustic and optical modes cannot be rigorously applied for non-crystals due to the lack of periodicity in the lattice, which restricts vibrations to non-propagating modes (e.g. diffusons and locons) \cite{phonons1, phonon2}. However, the phase quotient ($Q_p$) of Bell and Hibbins-Butler \cite{Bell_1975} provides a measure of how vibrations of neighboring atoms are in-phase (acoustic mode) and out-of-phase (optical mode). The normalized $Q_p$ is given as \cite{Allen_Feldman}:

\begin{figure}[!htpb]
	\includegraphics[width=.9\linewidth]{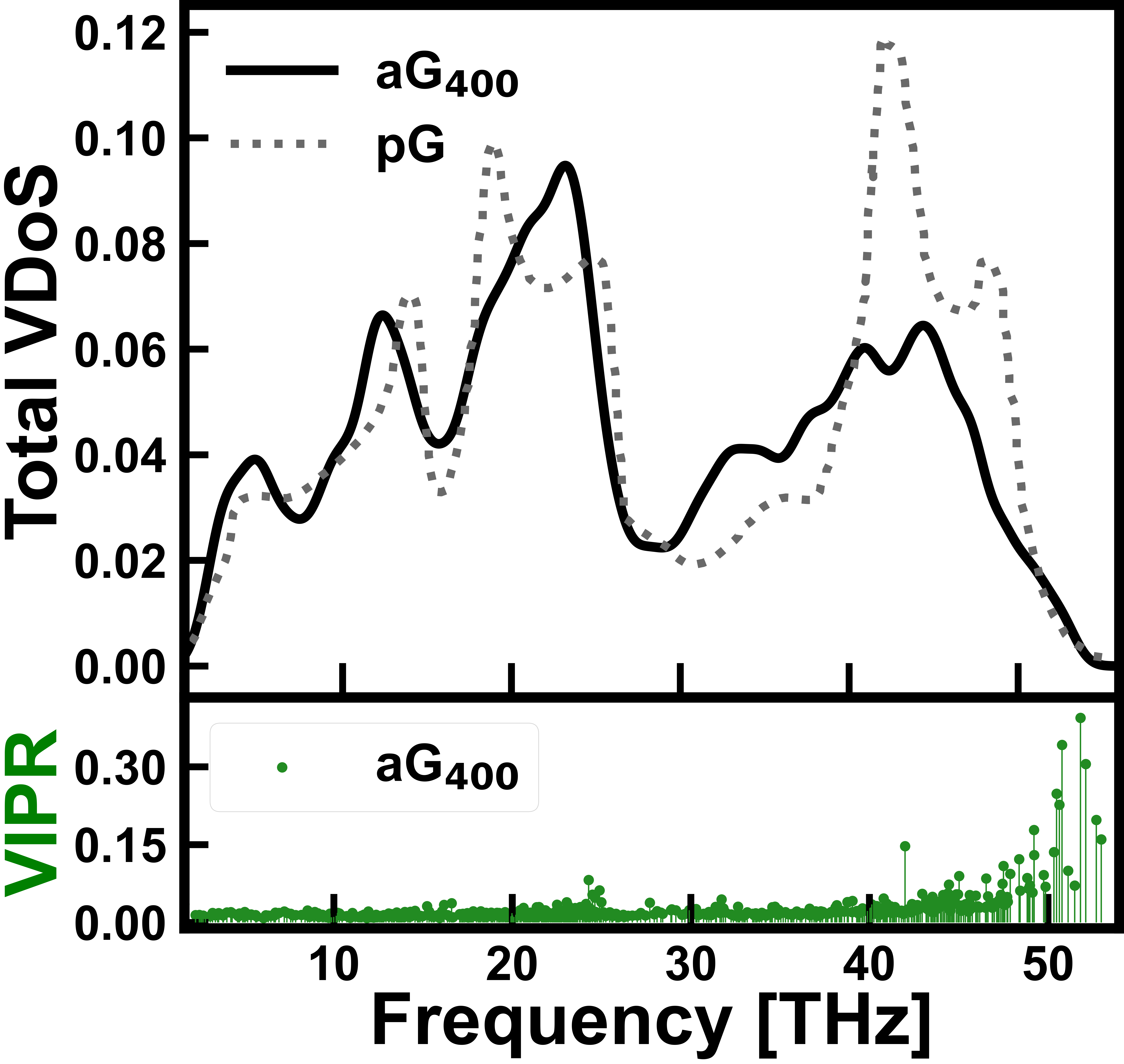}
	\caption{[TOP] The total VDoS for amorphous and pristine graphite calculated from the harmonic approximation as implemented within \texttt{VASP}. [BOTTOM]  VIPR amorphous Graphite. The result is shown for a 400-atom amorphous graphene model}
	\label{fig:Cfig_VDOS_VIPR}
\end{figure}

\begin{figure*}[!htpb]
	\centering
	\includegraphics[width=\textwidth]{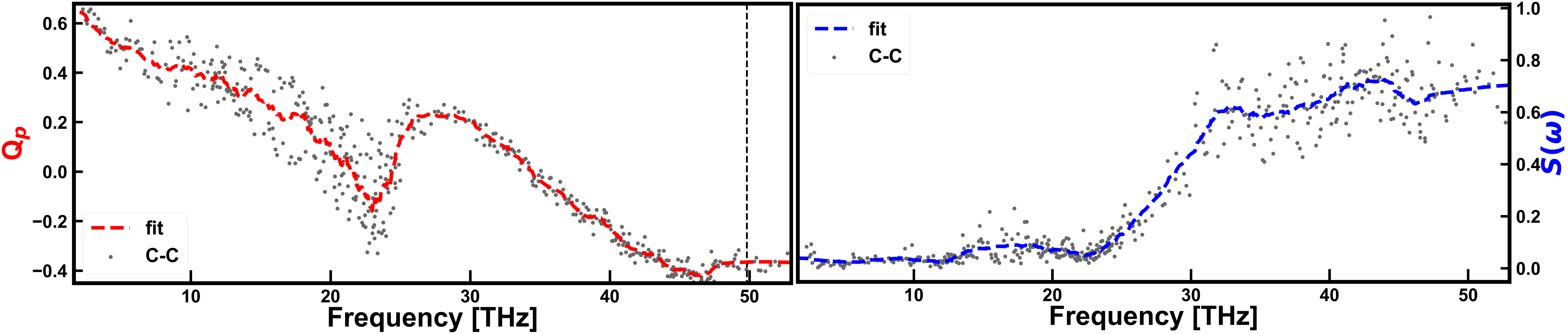}
    	\caption{Figure showing the [LEFT] phase quotient and [RIGHT] stretching character of aG. The dashed line shows the level at which VIPR = 0.15}
	\label{fig:Cfig_PQBS}
\end{figure*}

\begin{equation}
    Q_p ~= \frac{1}{N_b}\frac{\sum_{m} \boldsymbol{u}^{i}_{p} \cdot \boldsymbol{u}^{j}_{p}}{\sum_{m}| \boldsymbol{u}^{i}_{p} \cdot \boldsymbol{u}^{j}_{p}|}  
\end{equation}

\noindent where $N_b$ is the number of valance bonds, $\boldsymbol{u}^{i}_{p}$ and  $\boldsymbol{u}^{j}_{p}$ are the normalized displacement vectors (see Eq. \ref{eq:VIPR}) for the $p^{th}$ normal mode. The index, i, sums over all the C atoms and $j$ enumerates neighboring atoms of the $i^{th}$ atom. The vibration of the bulk material in unison gives $Q_p$ = 1 (purely acoustic). Conversely, a value of -1 would correspond to motion in the opposite direction between neighboring atoms (purely optical). It then follows that positive (negative) $Q_p$ is more ``acoustic-like” (``optical-like”). However near 0, one cannot necessarily distinguish between acoustic and optical modes. The phase quotient for aG$_{400}$ is plotted in Fig \ref{fig:Cfig_PQBS} [LEFT]. The dashed line identifies the diffuson to locon transition level, which is conceptually the region where VIPR $\gtrapprox$ 0.15 \cite{Henry35}. It then follows that the locons have high negative $Q_p$ values at the high-frequency end of the spectrum. Vibrational modes around the inflection point at $\approx$ 23 THz are not locons, but can be considered as quasi-localized ``Resonant modes". This results from the finite size of the supercell and diffuses away for larger systems \cite{Henry36}. The animation for vibration at one of the  Resonant mode frequencies (see quasiLocalized\_freq.mp4 in the supplementary material \cite{suppl}) confirms that the normal modes are not truly localized but rather majorly distributed amongst C atoms at the boundaries. 

We further investigated the vibration modes by calculating the bond-stretching character ($S(\omega)$) of aG using the following equation:

\begin{equation}
    S(\omega) ~= \frac{\sum_{m} |(\boldsymbol{u}^{i}_{n} - \boldsymbol{u}^{j}_{n}) \cdot \hat{\boldsymbol{r}}_{ij}|}{\sum_{m}| \boldsymbol{u}^{i}_{n} - \boldsymbol{u}^{j}_{n}|}  
\end{equation}

\noindent $\boldsymbol{u}^{i}_{n}$ and $\boldsymbol{u}^{i}_{n}$ are as defined in Eqn. \ref{eq:VIPR}, ~$\hat{\boldsymbol{r}}_{ij}$ is the unit vector parallel to the m$^{th}$ bond.  $S(\omega)$ is close to unity when the mode of vibration is predominantly of bond-stretching type and will be close to 0 otherwise. The Vibrations in aG were observed to be in-plane (out-of-plane) at high (low) frequencies, which are similar to what is found in pristine graphite \cite{vib1,vib2}. The in-plane vibrations involve the stretching of C-C bonds of neighboring atoms, while the out-of-plane vibrations correspond to the breathing mode  of individual layers in aG \cite{vib3,vib4}. Fig. \ref{fig:Cfig_PQBS} shows that bond stretching in the planes is dominant at high frequencies. We have included two animations in the supplementary material showing the stretching character and breathing mode at the extremes of the frequency spectrum \cite{suppl}. It is noteworthy that at the mid-spectrum (see quasiLocalized\_freq.mp4 \cite{suppl}), there is a combination of both characteristics as predicted from the phase quotient.
 
We point out that beyond the basic classification of vibrational modes in aG, the optical-like modes (negative $Q_p$ with the bond-stretching character) contribute to the thermal conductivity in disordered systems. This has been reported for amorphous carbon in the work of Hamid and coworkers \cite{Henry}, where they found that at high temperatures (with heat capacity satisfying the Dulong-Petit limit), vibration regions with negative $Q_p$ contribute as much as 40 \% to the total thermal conductivity. Using \texttt{LAMMPS}, we analyzed the contribution of the topological defect to the thermal conductivity (TC) in aG$_{540}$. The contribution of the heat flux (\textbf{J}) for each atom \cite{heatflux1, heatflux2} was calculated, and then an ensemble average of the auto-correlation of \textbf{J} was related to the TC ($\kappa$)  using the  Green-Kubo formalism given as \cite{Green, Kubo}:

\begin{equation}
    \kappa ~= \frac{1}{3VK_BT^2}\int_{0}^{\tau} \left < \textbf{J} (0) \cdot \textbf{J} (t) \right > dt
    \label{eq:kappa}
\end{equation}

where V, T, and $K_B$ are the system volume, temperature, and Boltzmann's constant, respectively. The upper limit of the integral was approximated by $\tau$ (= 0.5 ns) which is the correlation time required for the heat current autocorrelation to decay to zero. The TC was obtained by averaging the integral in Eq. \ref{eq:kappa} from 15 independent ensembles. The Nos\'e–Hoover thermostat \cite{nose, hoover} was used for thermalization and equilibration at T = 300 K and 1000 K in a fixed volume using a 1 fs time-step. At the beginning of the simulation, initial velocities were assigned to the atoms randomly from a Gaussian distribution. Our result showed that the average TC calculated for aG was 0.85 Wcm$^{-1}$K$^{-1}$ and 0.96 Wcm$^{-1}$K$^{-1}$ at 300 K and 1000 K respectively. The increase in the thermal conductivity is consistent with what is observed for amorphous systems like aC and amorphous silicon \cite{Henry, Allen_Feldman_aSi,TCofElements}. We note that the room temperature TC of aG is $\approx$ 5\% of that of pyrolytic graphite ($\kappa \approx$ 19.5 Wcm$^{-1}$K$^{-1}$) \cite{TCofElements}. However, the 11 \% increase for aG at 1000 K, compared to the $\approx$ 73 \% decrease for pyrolytic graphite \cite{TCofElements} at the same temperature, could be important for applications.

\section{\label{sec:conclusion} Conclusion}
This work focused on the structural, electronic, and vibration properties of amorphous graphite (aG). aG formed only with PBC in all dimensions within the density range of ca. 2.2-2.8 g/cm$^3$. However, at a "threshold" density of 2.0 g/cm$^3$, an undulating layered structure was observed. Structural features were analyzed and compared to pristine graphite by exploiting the radial distribution function and coordination number. Electronic structure analysis showed that there was no band-gap at the Fermi-level in aG. few states were observed to be localized on 5- and 7-member rings in the layers. Spatial projection of the charge density near the Fermi-level ($\pi$ orbitals) showed high values on connecting 6-member rings. The density of state and corresponding participation ratio for the phonon vibrations were analyzed, and the result showed that aG has more states localized at the high-frequency end of the vibration spectrum. phase quotient and stretching character analysis further suggested that those localized sites were from atoms participating in non-hexagonal rings. The average thermal conductivity for aG was calculated at room temperature (0.85 Wcm$^{-1}$K$^{-1}$) and 1000 K (0.96 Wcm$^{-1}$K$^{-1}$) indicated an 11 \% increase in thermal conductivity.

\begin{acknowledgments}
We thank the U.S. Department of Energy for support under Grant No. DE-FE0031981,  XSEDE (supported by National Science Foundation Grant No. ACI-1548562) for computational support under allocation no. DMR-190008P. 
\end{acknowledgments}

The Figures \ref{fig:Cfig_aGplanes} and \ref{fig:Kfig_localization} were made using the Open-Visualization Tool (\texttt{OVITO}) \cite{OVITO}. All the animations were made using \texttt{Jmol} \cite{Jmol}.

\bibliography{aG}

\providecommand{\noopsort}[1]{}\providecommand{\singleletter}[1]{#1}%
\begin{thebibliography}{57}%
\makeatletter
\providecommand \@ifxundefined [1]{%
 \@ifx{#1\undefined}
}%
\providecommand \@ifnum [1]{%
 \ifnum #1\expandafter \@firstoftwo
 \else \expandafter \@secondoftwo
 \fi
}%
\providecommand \@ifx [1]{%
 \ifx #1\expandafter \@firstoftwo
 \else \expandafter \@secondoftwo
 \fi
}%
\providecommand \natexlab [1]{#1}%
\providecommand \enquote  [1]{``#1''}%
\providecommand \bibnamefont  [1]{#1}%
\providecommand \bibfnamefont [1]{#1}%
\providecommand \citenamefont [1]{#1}%
\providecommand \href@noop [0]{\@secondoftwo}%
\providecommand \href [0]{\begingroup \@sanitize@url \@href}%
\providecommand \@href[1]{\@@startlink{#1}\@@href}%
\providecommand \@@href[1]{\endgroup#1\@@endlink}%
\providecommand \@sanitize@url [0]{\catcode `\\12\catcode `\$12\catcode
  `\&12\catcode `\#12\catcode `\^12\catcode `\_12\catcode `\%12\relax}%
\providecommand \@@startlink[1]{}%
\providecommand \@@endlink[0]{}%
\providecommand \url  [0]{\begingroup\@sanitize@url \@url }%
\providecommand \@url [1]{\endgroup\@href {#1}{\urlprefix }}%
\providecommand \urlprefix  [0]{URL }%
\providecommand \Eprint [0]{\href }%
\providecommand \doibase [0]{https://doi.org/}%
\providecommand \selectlanguage [0]{\@gobble}%
\providecommand \bibinfo  [0]{\@secondoftwo}%
\providecommand \bibfield  [0]{\@secondoftwo}%
\providecommand \translation [1]{[#1]}%
\providecommand \BibitemOpen [0]{}%
\providecommand \bibitemStop [0]{}%
\providecommand \bibitemNoStop [0]{.\EOS\space}%
\providecommand \EOS [0]{\spacefactor3000\relax}%
\providecommand \BibitemShut  [1]{\csname bibitem#1\endcsname}%
\let\auto@bib@innerbib\@empty
\bibitem [{\citenamefont {Olson}\ \emph {et~al.}(2016)\citenamefont {Olson},
  \citenamefont {Virta}, \citenamefont {Mahdavi}, \citenamefont {Sangine},\
  and\ \citenamefont {Fortier}}]{flakesbook}%
  \BibitemOpen
  \bibfield  {author} {\bibinfo {author} {\bibfnamefont {D.}~\bibnamefont
  {Olson}}, \bibinfo {author} {\bibfnamefont {R.}~\bibnamefont {Virta}},
  \bibinfo {author} {\bibfnamefont {M.}~\bibnamefont {Mahdavi}}, \bibinfo
  {author} {\bibfnamefont {E.}~\bibnamefont {Sangine}},\ and\ \bibinfo {author}
  {\bibfnamefont {S.}~\bibnamefont {Fortier}},\ }\bibfield  {title} {\bibinfo
  {title} {Natural graphite demand and supply--implications for electric
  vehicle battery requirements},\ }\href
  {https://doi.org/10.1130/2016.2520(08)} {\bibfield  {journal} {\bibinfo
  {journal} {The Geological Society of America}\ }\textbf {\bibinfo {volume}
  {520}},\ \bibinfo {pages} {67} (\bibinfo {year} {2016})}\BibitemShut
  {NoStop}%
\bibitem [{\citenamefont {Mills}()}]{flakes1}%
  \BibitemOpen
  \bibfield  {author} {\bibinfo {author} {\bibfnamefont {R.}~\bibnamefont
  {Mills}},\ }\bibfield  {title} {\bibinfo {title} {{Graphite deficit starting
  this year, as demand for EV battery anode ingredient exceeds supply}},\
  }\href
  {https://www.mining.com/web/graphite-deficit-starting-this-year-as-demand-for-ev-battery-anode-ingredient-exceeds-supply}
  {\bibfield  {journal} {\bibinfo  {journal}
  {https://www.mining.com/web/graphite-deficit-starting-this-year-as-demand-for-ev-battery-anode-ingredient-exceeds-supply}\
  }}\bibinfo {note} {Accessed: 2022-08-03}\BibitemShut {NoStop}%
\bibitem [{USG()}]{USGS}%
  \BibitemOpen
  \href@noop {} {\bibinfo {title} {{U.S. Geological Survey, Mineral Commodity
  Summaries, 2022-Graphite}}},\ \bibinfo {howpublished}
  {\url{https://pubs.usgs.gov/periodicals/mcs2022/mcs2022-graphite.pdf}},\
  \bibinfo {note} {accessed: 2022-06-25}\BibitemShut {NoStop}%
\bibitem [{\citenamefont {Gao}\ \emph {et~al.}(2020)\citenamefont {Gao},
  \citenamefont {Wang}, \citenamefont {Zhang}, \citenamefont {Jing},
  \citenamefont {Ma}, \citenamefont {Chen},\ and\ \citenamefont {Zhang}}]{R1}%
  \BibitemOpen
  \bibfield  {author} {\bibinfo {author} {\bibfnamefont {Y.}~\bibnamefont
  {Gao}}, \bibinfo {author} {\bibfnamefont {C.}~\bibnamefont {Wang}}, \bibinfo
  {author} {\bibfnamefont {J.}~\bibnamefont {Zhang}}, \bibinfo {author}
  {\bibfnamefont {Q.}~\bibnamefont {Jing}}, \bibinfo {author} {\bibfnamefont
  {B.}~\bibnamefont {Ma}}, \bibinfo {author} {\bibfnamefont {Y.}~\bibnamefont
  {Chen}},\ and\ \bibinfo {author} {\bibfnamefont {W.}~\bibnamefont {Zhang}},\
  }\bibfield  {title} {\bibinfo {title} {Graphite recycling from the spent
  lithium-ion batteries by sulfuric acid curing–leaching combined with
  high-temperature calcination},\ }\href
  {https://doi.org/10.1021/acssuschemeng.0c02321} {\bibfield  {journal}
  {\bibinfo  {journal} {ACS Sustainable Chemistry \& Engineering}\ }\textbf
  {\bibinfo {volume} {8}},\ \bibinfo {pages} {9447} (\bibinfo {year}
  {2020})}\BibitemShut {NoStop}%
\bibitem [{\citenamefont {Liu}\ \emph {et~al.}(2022)\citenamefont {Liu},
  \citenamefont {Qu}, \citenamefont {Zhang}, \citenamefont {Zhao},
  \citenamefont {Xie},\ and\ \citenamefont {Yin}}]{R2}%
  \BibitemOpen
  \bibfield  {author} {\bibinfo {author} {\bibfnamefont {D.}~\bibnamefont
  {Liu}}, \bibinfo {author} {\bibfnamefont {X.}~\bibnamefont {Qu}}, \bibinfo
  {author} {\bibfnamefont {B.}~\bibnamefont {Zhang}}, \bibinfo {author}
  {\bibfnamefont {J.}~\bibnamefont {Zhao}}, \bibinfo {author} {\bibfnamefont
  {H.}~\bibnamefont {Xie}},\ and\ \bibinfo {author} {\bibfnamefont
  {H.}~\bibnamefont {Yin}},\ }\bibfield  {title} {\bibinfo {title} {Alkaline
  roasting approach to reclaiming lithium and graphite from spent lithium-ion
  batteries},\ }\href {https://doi.org/10.1021/acssuschemeng.1c07852}
  {\bibfield  {journal} {\bibinfo  {journal} {ACS Sustainable Chemistry \&
  Engineering}\ }\textbf {\bibinfo {volume} {10}},\ \bibinfo {pages} {5739}
  (\bibinfo {year} {2022})}\BibitemShut {NoStop}%
\bibitem [{\citenamefont {Bhar}\ \emph {et~al.}(2022)\citenamefont {Bhar},
  \citenamefont {Ghosh}, \citenamefont {Krishnamurthy}, \citenamefont
  {Yalamanchili},\ and\ \citenamefont {Martha}}]{R3}%
  \BibitemOpen
  \bibfield  {author} {\bibinfo {author} {\bibfnamefont {M.}~\bibnamefont
  {Bhar}}, \bibinfo {author} {\bibfnamefont {S.}~\bibnamefont {Ghosh}},
  \bibinfo {author} {\bibfnamefont {S.}~\bibnamefont {Krishnamurthy}}, \bibinfo
  {author} {\bibfnamefont {K.}~\bibnamefont {Yalamanchili}},\ and\ \bibinfo
  {author} {\bibfnamefont {S.~K.}\ \bibnamefont {Martha}},\ }\bibfield  {title}
  {\bibinfo {title} {Electrochemical compatibility of graphite anode from spent
  li-ion batteries: Recycled via a greener and sustainable approach},\ }\href
  {https://doi.org/10.1021/acssuschemeng.2c00554} {\bibfield  {journal}
  {\bibinfo  {journal} {ACS Sustainable Chemistry \& Engineering}\ }\textbf
  {\bibinfo {volume} {10}},\ \bibinfo {pages} {7515} (\bibinfo {year}
  {2022})}\BibitemShut {NoStop}%
\bibitem [{\citenamefont {Natarajan}\ \emph {et~al.}(2022)\citenamefont
  {Natarajan}, \citenamefont {Krishnamoorthy}, \citenamefont {Sathyaseelan},
  \citenamefont {Mariappan}, \citenamefont {Pazhamalai}, \citenamefont
  {Manoharan},\ and\ \citenamefont {Kim}}]{R4}%
  \BibitemOpen
  \bibfield  {author} {\bibinfo {author} {\bibfnamefont {S.}~\bibnamefont
  {Natarajan}}, \bibinfo {author} {\bibfnamefont {K.}~\bibnamefont
  {Krishnamoorthy}}, \bibinfo {author} {\bibfnamefont {A.}~\bibnamefont
  {Sathyaseelan}}, \bibinfo {author} {\bibfnamefont {V.~K.}\ \bibnamefont
  {Mariappan}}, \bibinfo {author} {\bibfnamefont {P.}~\bibnamefont
  {Pazhamalai}}, \bibinfo {author} {\bibfnamefont {S.}~\bibnamefont
  {Manoharan}},\ and\ \bibinfo {author} {\bibfnamefont {S.-J.}\ \bibnamefont
  {Kim}},\ }\bibfield  {title} {\bibinfo {title} {A new route for the recycling
  of spent lithium-ion batteries towards advanced energy storage, conversion,
  and harvesting systems},\ }\href
  {https://doi.org/https://doi.org/10.1016/j.nanoen.2022.107595} {\bibfield
  {journal} {\bibinfo  {journal} {Nano Energy}\ }\textbf {\bibinfo {volume}
  {101}},\ \bibinfo {pages} {107595} (\bibinfo {year} {2022})}\BibitemShut
  {NoStop}%
\bibitem [{\citenamefont {Yi}\ \emph {et~al.}(2022)\citenamefont {Yi},
  \citenamefont {Ge}, \citenamefont {Wu}, \citenamefont {Sun},\ and\
  \citenamefont {Yang}}]{R5}%
  \BibitemOpen
  \bibfield  {author} {\bibinfo {author} {\bibfnamefont {C.}~\bibnamefont
  {Yi}}, \bibinfo {author} {\bibfnamefont {P.}~\bibnamefont {Ge}}, \bibinfo
  {author} {\bibfnamefont {X.}~\bibnamefont {Wu}}, \bibinfo {author}
  {\bibfnamefont {W.}~\bibnamefont {Sun}},\ and\ \bibinfo {author}
  {\bibfnamefont {Y.}~\bibnamefont {Yang}},\ }\bibfield  {title} {\bibinfo
  {title} {Tailoring carbon chains for repairing graphite from spent
  lithium-ion battery toward closed-circuit recycling},\ }\href
  {https://doi.org/https://doi.org/10.1016/j.jechem.2022.05.002} {\bibfield
  {journal} {\bibinfo  {journal} {Journal of Energy Chemistry}\ }\textbf
  {\bibinfo {volume} {72}},\ \bibinfo {pages} {97} (\bibinfo {year}
  {2022})}\BibitemShut {NoStop}%
\bibitem [{\citenamefont {Velenturf}\ and\ \citenamefont
  {Purnell}(2021)}]{SL0}%
  \BibitemOpen
  \bibfield  {author} {\bibinfo {author} {\bibfnamefont {A.~P.}\ \bibnamefont
  {Velenturf}}\ and\ \bibinfo {author} {\bibfnamefont {P.}~\bibnamefont
  {Purnell}},\ }\bibfield  {title} {\bibinfo {title} {Principles for a
  sustainable circular economy},\ }\href
  {https://doi.org/https://doi.org/10.1016/j.spc.2021.02.018} {\bibfield
  {journal} {\bibinfo  {journal} {Sustainable Production and Consumption}\
  }\textbf {\bibinfo {volume} {27}},\ \bibinfo {pages} {1437} (\bibinfo {year}
  {2021})}\BibitemShut {NoStop}%
\bibitem [{\citenamefont {Rey}\ \emph {et~al.}(2021)\citenamefont {Rey},
  \citenamefont {Vallejo}, \citenamefont {Santiago}, \citenamefont
  {Iturrondobeitia},\ and\ \citenamefont {Lizundia}}]{SL1}%
  \BibitemOpen
  \bibfield  {author} {\bibinfo {author} {\bibfnamefont {I.}~\bibnamefont
  {Rey}}, \bibinfo {author} {\bibfnamefont {C.}~\bibnamefont {Vallejo}},
  \bibinfo {author} {\bibfnamefont {G.}~\bibnamefont {Santiago}}, \bibinfo
  {author} {\bibfnamefont {M.}~\bibnamefont {Iturrondobeitia}},\ and\ \bibinfo
  {author} {\bibfnamefont {E.}~\bibnamefont {Lizundia}},\ }\bibfield  {title}
  {\bibinfo {title} {Environmental impacts of graphite recycling from spent
  lithium-ion batteries based on life cycle assessment},\ }\href
  {https://doi.org/10.1021/acssuschemeng.1c04938} {\bibfield  {journal}
  {\bibinfo  {journal} {ACS Sustainable Chemistry \& Engineering}\ }\textbf
  {\bibinfo {volume} {9}},\ \bibinfo {pages} {14488} (\bibinfo {year}
  {2021})}\BibitemShut {NoStop}%
\bibitem [{\citenamefont {Niese}\ \emph {et~al.}()\citenamefont {Niese},
  \citenamefont {Pieper}, \citenamefont {Arora}, ,\ and\ \citenamefont
  {Xie}}]{SL2}%
  \BibitemOpen
  \bibfield  {author} {\bibinfo {author} {\bibfnamefont {N.}~\bibnamefont
  {Niese}}, \bibinfo {author} {\bibfnamefont {C.}~\bibnamefont {Pieper}},
  \bibinfo {author} {\bibfnamefont {A.}~\bibnamefont {Arora}}, ,\ and\ \bibinfo
  {author} {\bibfnamefont {A.}~\bibnamefont {Xie}},\ }\bibfield  {title}
  {\bibinfo {title} {{The Case for a Circular Economy in Electric Vehicle
  Batteries}},\ }\href
  {https://www.bcg.com/publications/2020/case-for-circular-economy-in-electric-vehicle-batteries}
  {\bibfield  {journal} {\bibinfo  {journal}
  {https://www.bcg.com/publications/2020/case-for-circular-economy-in-electric-vehicle-batteries}\
  }}\bibinfo {note} {Accessed: 2022-08-07}\BibitemShut {NoStop}%
\bibitem [{\citenamefont {{Pham, Thi Thuy Linh}}(2021)}]{SL3}%
  \BibitemOpen
  \bibfield  {author} {\bibinfo {author} {\bibnamefont {{Pham, Thi Thuy
  Linh}}},\ }\href@noop {} {\bibinfo {title} {{The second life - Challenges of
  repurposing electric vehicle lithium-ion batteries}}} (\bibinfo {year}
  {{2021}}),\ \bibinfo {note} {{The International Institute for Industrial
  Environmental Economics Master Thesis}}\BibitemShut {NoStop}%
\bibitem [{\citenamefont {Masi}\ \emph {et~al.}(2021)\citenamefont {Masi},
  \citenamefont {Schumacher}, \citenamefont {Hilman}, \citenamefont {Dulal},
  \citenamefont {Rimal}, \citenamefont {Xu}, \citenamefont {Leonard},
  \citenamefont {Tang}, \citenamefont {Fan},\ and\ \citenamefont
  {Chien}}]{gp1}%
  \BibitemOpen
  \bibfield  {author} {\bibinfo {author} {\bibfnamefont {C.~A.}\ \bibnamefont
  {Masi}}, \bibinfo {author} {\bibfnamefont {T.~A.}\ \bibnamefont
  {Schumacher}}, \bibinfo {author} {\bibfnamefont {J.}~\bibnamefont {Hilman}},
  \bibinfo {author} {\bibfnamefont {R.}~\bibnamefont {Dulal}}, \bibinfo
  {author} {\bibfnamefont {G.}~\bibnamefont {Rimal}}, \bibinfo {author}
  {\bibfnamefont {B.}~\bibnamefont {Xu}}, \bibinfo {author} {\bibfnamefont
  {B.}~\bibnamefont {Leonard}}, \bibinfo {author} {\bibfnamefont
  {J.}~\bibnamefont {Tang}}, \bibinfo {author} {\bibfnamefont {M.}~\bibnamefont
  {Fan}},\ and\ \bibinfo {author} {\bibfnamefont {T.}~\bibnamefont {Chien}},\
  }\bibfield  {title} {\bibinfo {title} {Converting raw coal powder into
  polycrystalline nano-graphite by metal-assisted microwave treatment},\ }\href
  {https://doi.org/https://doi.org/10.1016/j.nanoso.2020.100660} {\bibfield
  {journal} {\bibinfo  {journal} {Nano-Structures and Nano-Objects}\ }\textbf
  {\bibinfo {volume} {25}},\ \bibinfo {pages} {100660} (\bibinfo {year}
  {2021})}\BibitemShut {NoStop}%
\bibitem [{\citenamefont {Qiu}\ \emph {et~al.}(2020)\citenamefont {Qiu},
  \citenamefont {Yang},\ and\ \citenamefont {Bai}}]{gp2}%
  \BibitemOpen
  \bibfield  {author} {\bibinfo {author} {\bibfnamefont {T.}~\bibnamefont
  {Qiu}}, \bibinfo {author} {\bibfnamefont {J.-G.}\ \bibnamefont {Yang}},\ and\
  \bibinfo {author} {\bibfnamefont {X.-J.}\ \bibnamefont {Bai}},\ }\bibfield
  {title} {\bibinfo {title} {Preparation of coal-based graphite with different
  microstructures by adjusting the content of ash and volatile matter in raw
  coal},\ }\href {https://doi.org/10.1080/15567036.2019.1604900} {\bibfield
  {journal} {\bibinfo  {journal} {Energy Sources, Part A: Recovery,
  Utilization, and Environmental Effects}\ }\textbf {\bibinfo {volume} {42}},\
  \bibinfo {pages} {1874} (\bibinfo {year} {2020})},\ \Eprint
  {https://arxiv.org/abs/https://doi.org/10.1080/15567036.2019.1604900}
  {https://doi.org/10.1080/15567036.2019.1604900} \BibitemShut {NoStop}%
\bibitem [{\citenamefont {Marsh}\ and\ \citenamefont
  {Rodríguez-Reinoso}(2006)}]{gp3}%
  \BibitemOpen
  \bibfield  {author} {\bibinfo {author} {\bibfnamefont {H.}~\bibnamefont
  {Marsh}}\ and\ \bibinfo {author} {\bibfnamefont {F.}~\bibnamefont
  {Rodríguez-Reinoso}},\ }\bibfield  {title} {\bibinfo {title} {Chapter 9 -
  production and reference material},\ }in\ \href
  {https://doi.org/https://doi.org/10.1016/B978-008044463-5/50023-6} {\emph
  {\bibinfo {booktitle} {Activated Carbon}}},\ \bibinfo {editor} {edited by\
  \bibinfo {editor} {\bibfnamefont {H.}~\bibnamefont {Marsh}}\ and\ \bibinfo
  {editor} {\bibfnamefont {F.}~\bibnamefont {Rodríguez-Reinoso}}}\ (\bibinfo
  {publisher} {Elsevier Science Ltd},\ \bibinfo {address} {Oxford},\ \bibinfo
  {year} {2006})\ pp.\ \bibinfo {pages} {454--508}\BibitemShut {NoStop}%
\bibitem [{\citenamefont {Qiu}\ \emph {et~al.}(2022)\citenamefont {Qiu},
  \citenamefont {Yu}, \citenamefont {Xie}, \citenamefont {He}, \citenamefont
  {Wang},\ and\ \citenamefont {Zhang}}]{gp4}%
  \BibitemOpen
  \bibfield  {author} {\bibinfo {author} {\bibfnamefont {T.}~\bibnamefont
  {Qiu}}, \bibinfo {author} {\bibfnamefont {Z.}~\bibnamefont {Yu}}, \bibinfo
  {author} {\bibfnamefont {W.}~\bibnamefont {Xie}}, \bibinfo {author}
  {\bibfnamefont {Y.}~\bibnamefont {He}}, \bibinfo {author} {\bibfnamefont
  {H.}~\bibnamefont {Wang}},\ and\ \bibinfo {author} {\bibfnamefont
  {T.}~\bibnamefont {Zhang}},\ }\bibfield  {title} {\bibinfo {title}
  {Preparation of onion-like synthetic graphite with a hierarchical pore
  structure from anthracite and its electrochemical properties as the anode
  material of lithium-ion batteries},\ }\href
  {https://doi.org/10.1021/acs.energyfuels.2c01892} {\bibfield  {journal}
  {\bibinfo  {journal} {Energy \& Fuels}\ }\textbf {\bibinfo {volume} {36}},\
  \bibinfo {pages} {8256} (\bibinfo {year} {2022})},\ \Eprint
  {https://arxiv.org/abs/https://doi.org/10.1021/acs.energyfuels.2c01892}
  {https://doi.org/10.1021/acs.energyfuels.2c01892} \BibitemShut {NoStop}%
\bibitem [{\citenamefont {Adamczyk}\ \emph {et~al.}(2021)\citenamefont
  {Adamczyk}, \citenamefont {Komorek}, \citenamefont {Białecka}, \citenamefont
  {Moszko},\ and\ \citenamefont {Klupa}}]{gp5}%
  \BibitemOpen
  \bibfield  {author} {\bibinfo {author} {\bibfnamefont {Z.}~\bibnamefont
  {Adamczyk}}, \bibinfo {author} {\bibfnamefont {J.}~\bibnamefont {Komorek}},
  \bibinfo {author} {\bibfnamefont {B.}~\bibnamefont {Białecka}}, \bibinfo
  {author} {\bibfnamefont {J.}~\bibnamefont {Moszko}},\ and\ \bibinfo {author}
  {\bibfnamefont {A.}~\bibnamefont {Klupa}},\ }\bibfield  {title} {\bibinfo
  {title} {Possibilities of graphitization of unburned carbon from coal fly
  ash},\ }\href {https://doi.org/10.3390/min11091027} {\bibfield  {journal}
  {\bibinfo  {journal} {Minerals}\ }\textbf {\bibinfo {volume} {11}},\ \bibinfo
  {pages} {1027} (\bibinfo {year} {2021})}\BibitemShut {NoStop}%
\bibitem [{\citenamefont {Wu}\ \emph {et~al.}(2021)\citenamefont {Wu},
  \citenamefont {Li}, \citenamefont {Wang}, \citenamefont {Hu}, \citenamefont
  {Cao},\ and\ \citenamefont {Liu}}]{gp6}%
  \BibitemOpen
  \bibfield  {author} {\bibinfo {author} {\bibfnamefont {Y.}~\bibnamefont
  {Wu}}, \bibinfo {author} {\bibfnamefont {K.}~\bibnamefont {Li}}, \bibinfo
  {author} {\bibfnamefont {Z.}~\bibnamefont {Wang}}, \bibinfo {author}
  {\bibfnamefont {M.}~\bibnamefont {Hu}}, \bibinfo {author} {\bibfnamefont
  {H.}~\bibnamefont {Cao}},\ and\ \bibinfo {author} {\bibfnamefont
  {Q.}~\bibnamefont {Liu}},\ }\bibfield  {title} {\bibinfo {title}
  {Fluctuations in graphitization of coal seam-derived natural graphite upon
  approaching the qitianling granite intrusion, hunan, china},\ }\bibfield
  {journal} {\bibinfo  {journal} {Minerals}\ }\textbf {\bibinfo {volume}
  {11}},\ \href {https://doi.org/10.3390/min11101147} {10.3390/min11101147}
  (\bibinfo {year} {2021})\BibitemShut {NoStop}%
\bibitem [{\citenamefont {Thapa}\ \emph {et~al.}(2022)\citenamefont {Thapa},
  \citenamefont {Ugwumadu}, \citenamefont {Nepal}, \citenamefont {Trembly},\
  and\ \citenamefont {Drabold}}]{LAG}%
  \BibitemOpen
  \bibfield  {author} {\bibinfo {author} {\bibfnamefont {R.}~\bibnamefont
  {Thapa}}, \bibinfo {author} {\bibfnamefont {C.}~\bibnamefont {Ugwumadu}},
  \bibinfo {author} {\bibfnamefont {K.}~\bibnamefont {Nepal}}, \bibinfo
  {author} {\bibfnamefont {J.}~\bibnamefont {Trembly}},\ and\ \bibinfo {author}
  {\bibfnamefont {D.~A.}\ \bibnamefont {Drabold}},\ }\bibfield  {title}
  {\bibinfo {title} {Ab initio simulation of amorphous graphite},\ }\href
  {https://doi.org/10.1103/PhysRevLett.128.236402} {\bibfield  {journal}
  {\bibinfo  {journal} {Phys. Rev. Lett.}\ }\textbf {\bibinfo {volume} {128}},\
  \bibinfo {pages} {236402} (\bibinfo {year} {2022})}\BibitemShut {NoStop}%
\bibitem [{\citenamefont {Wyckoff}(1963)}]{pG_EXAFS}%
  \BibitemOpen
  \bibfield  {author} {\bibinfo {author} {\bibfnamefont {R.}~\bibnamefont
  {Wyckoff}},\ }\href {https://books.google.com/books?id=PttqPAUNgeUC} {\emph
  {\bibinfo {title} {Crystal Structures}}},\ \bibinfo {series} {Crystal
  Structures}\ No.\ \bibinfo {number} {v. 1}\ (\bibinfo  {publisher}
  {Interscience Publishers},\ \bibinfo {year} {1963})\BibitemShut {NoStop}%
\bibitem [{\citenamefont {Bhattarai}\ \emph {et~al.}(2018)\citenamefont
  {Bhattarai}, \citenamefont {Biswas}, \citenamefont {Atta-Fynn},\ and\
  \citenamefont {Drabold}}]{DAD_aC}%
  \BibitemOpen
  \bibfield  {author} {\bibinfo {author} {\bibfnamefont {B.}~\bibnamefont
  {Bhattarai}}, \bibinfo {author} {\bibfnamefont {P.}~\bibnamefont {Biswas}},
  \bibinfo {author} {\bibfnamefont {R.}~\bibnamefont {Atta-Fynn}},\ and\
  \bibinfo {author} {\bibfnamefont {D.~A.}\ \bibnamefont {Drabold}},\
  }\bibfield  {title} {\bibinfo {title} {Amorphous graphene: a constituent part
  of low density amorphous carbon},\ }\href
  {https://doi.org/10.1039/C8CP02545B} {\bibfield  {journal} {\bibinfo
  {journal} {Phys. Chem. Chem. Phys.}\ }\textbf {\bibinfo {volume} {20}},\
  \bibinfo {pages} {19546} (\bibinfo {year} {2018})}\BibitemShut {NoStop}%
\bibitem [{\citenamefont {Kresse}\ and\ \citenamefont
  {Furthm\"uller}(1996)}]{VASP}%
  \BibitemOpen
  \bibfield  {author} {\bibinfo {author} {\bibfnamefont {G.}~\bibnamefont
  {Kresse}}\ and\ \bibinfo {author} {\bibfnamefont {J.}~\bibnamefont
  {Furthm\"uller}},\ }\bibfield  {title} {\bibinfo {title} {Efficient iterative
  schemes for ab initio total-energy calculations using a plane-wave basis
  set},\ }\href {https://doi.org/10.1103/PhysRevB.54.11169} {\bibfield
  {journal} {\bibinfo  {journal} {Phys. Rev. B}\ }\textbf {\bibinfo {volume}
  {54}},\ \bibinfo {pages} {11169} (\bibinfo {year} {1996})}\BibitemShut
  {NoStop}%
\bibitem [{\citenamefont {Thompson}\ \emph {et~al.}(2022)\citenamefont
  {Thompson}, \citenamefont {Aktulga}, \citenamefont {Berger}, \citenamefont
  {Bolintineanu}, \citenamefont {Brown}, \citenamefont {Crozier}, \citenamefont
  {{in 't Veld}}, \citenamefont {Kohlmeyer}, \citenamefont {Moore},
  \citenamefont {Nguyen}, \citenamefont {Shan}, \citenamefont {Stevens},
  \citenamefont {Tranchida}, \citenamefont {Trott},\ and\ \citenamefont
  {Plimpton}}]{lammps}%
  \BibitemOpen
  \bibfield  {author} {\bibinfo {author} {\bibfnamefont {A.~P.}\ \bibnamefont
  {Thompson}}, \bibinfo {author} {\bibfnamefont {H.~M.}\ \bibnamefont
  {Aktulga}}, \bibinfo {author} {\bibfnamefont {R.}~\bibnamefont {Berger}},
  \bibinfo {author} {\bibfnamefont {D.~S.}\ \bibnamefont {Bolintineanu}},
  \bibinfo {author} {\bibfnamefont {W.~M.}\ \bibnamefont {Brown}}, \bibinfo
  {author} {\bibfnamefont {P.~S.}\ \bibnamefont {Crozier}}, \bibinfo {author}
  {\bibfnamefont {P.~J.}\ \bibnamefont {{in 't Veld}}}, \bibinfo {author}
  {\bibfnamefont {A.}~\bibnamefont {Kohlmeyer}}, \bibinfo {author}
  {\bibfnamefont {S.~G.}\ \bibnamefont {Moore}}, \bibinfo {author}
  {\bibfnamefont {T.~D.}\ \bibnamefont {Nguyen}}, \bibinfo {author}
  {\bibfnamefont {R.}~\bibnamefont {Shan}}, \bibinfo {author} {\bibfnamefont
  {M.~J.}\ \bibnamefont {Stevens}}, \bibinfo {author} {\bibfnamefont
  {J.}~\bibnamefont {Tranchida}}, \bibinfo {author} {\bibfnamefont
  {C.}~\bibnamefont {Trott}},\ and\ \bibinfo {author} {\bibfnamefont {S.~J.}\
  \bibnamefont {Plimpton}},\ }\bibfield  {title} {\bibinfo {title} {{LAMMPS - a
  flexible simulation tool for particle-based materials modeling at the atomic,
  meso, and continuum scales}},\ }\href
  {https://doi.org/https://doi.org/10.1016/j.cpc.2021.108171} {\bibfield
  {journal} {\bibinfo  {journal} {Computer Physics Communications}\ }\textbf
  {\bibinfo {volume} {271}},\ \bibinfo {pages} {108171} (\bibinfo {year}
  {2022})}\BibitemShut {NoStop}%
\bibitem [{\citenamefont {Deringer}\ and\ \citenamefont {Cs\'anyi}(2017)}]{C}%
  \BibitemOpen
  \bibfield  {author} {\bibinfo {author} {\bibfnamefont {V.~L.}\ \bibnamefont
  {Deringer}}\ and\ \bibinfo {author} {\bibfnamefont {G.}~\bibnamefont
  {Cs\'anyi}},\ }\bibfield  {title} {\bibinfo {title} {Machine learning based
  interatomic potential for amorphous carbon},\ }\href
  {https://doi.org/10.1103/PhysRevB.95.094203} {\bibfield  {journal} {\bibinfo
  {journal} {Phys. Rev. B}\ }\textbf {\bibinfo {volume} {95}},\ \bibinfo
  {pages} {094203} (\bibinfo {year} {2017})}\BibitemShut {NoStop}%
\bibitem [{sup()}]{suppl}%
  \BibitemOpen
  \href@noop {} {\bibinfo {title} {{Supplemental materials with animation
  showing the formation of aG and its different phonon vibration
  modes}}}\BibitemShut {NoStop}%
\bibitem [{\citenamefont {Ugwumadu}\ \emph {et~al.}()\citenamefont {Ugwumadu},
  \citenamefont {Nepal}, \citenamefont {Thapa}, \citenamefont {Lee},
  \citenamefont {Majali}, \citenamefont {Trembly},\ and\ \citenamefont
  {Drabold}}]{BO}%
  \BibitemOpen
  \bibfield  {author} {\bibinfo {author} {\bibfnamefont {C.}~\bibnamefont
  {Ugwumadu}}, \bibinfo {author} {\bibfnamefont {K.}~\bibnamefont {Nepal}},
  \bibinfo {author} {\bibfnamefont {R.}~\bibnamefont {Thapa}}, \bibinfo
  {author} {\bibfnamefont {Y.~G.}\ \bibnamefont {Lee}}, \bibinfo {author}
  {\bibfnamefont {Y.~A.}\ \bibnamefont {Majali}}, \bibinfo {author}
  {\bibfnamefont {J.}~\bibnamefont {Trembly}},\ and\ \bibinfo {author}
  {\bibfnamefont {D.~A.}\ \bibnamefont {Drabold}},\ }\bibfield  {title}
  {\bibinfo {title} {{Simulation of multi-shell fullerenes using
  machine-learning Gaussian Approximation Potential}},\ }\bibinfo {note}
  {submitted to Carbon Trends on 08/2022. Available at SSRN:
  \url{http://dx.doi.org/10.2139/ssrn.4200272}}\BibitemShut {NoStop}%
\bibitem [{\citenamefont {Mackay}\ and\ \citenamefont
  {Terrones}(1991)}]{Mackay}%
  \BibitemOpen
  \bibfield  {author} {\bibinfo {author} {\bibfnamefont {A.~L.}\ \bibnamefont
  {Mackay}}\ and\ \bibinfo {author} {\bibfnamefont {H.}~\bibnamefont
  {Terrones}},\ }\bibfield  {title} {\bibinfo {title} {Diamond from graphite},\
  }\href {https://doi.org/10.1038/352762a0} {\bibfield  {journal} {\bibinfo
  {journal} {Nature}\ }\textbf {\bibinfo {volume} {352}},\ \bibinfo {pages}
  {762} (\bibinfo {year} {1991})}\BibitemShut {NoStop}%
\bibitem [{\citenamefont {Terrones}\ and\ \citenamefont
  {Mackay}(1993)}]{TERRONES}%
  \BibitemOpen
  \bibfield  {author} {\bibinfo {author} {\bibfnamefont {H.}~\bibnamefont
  {Terrones}}\ and\ \bibinfo {author} {\bibfnamefont {A.}~\bibnamefont
  {Mackay}},\ }\bibfield  {title} {\bibinfo {title} {Hypothetical curved
  graphite},\ }\href
  {https://doi.org/https://doi.org/10.1016/0965-9773(93)90094-R} {\bibfield
  {journal} {\bibinfo  {journal} {Nanostructured Materials}\ }\textbf {\bibinfo
  {volume} {3}},\ \bibinfo {pages} {319} (\bibinfo {year} {1993})},\ \bibinfo
  {note} {proceedings of the First International Conference on Nanostructured
  Materials}\BibitemShut {NoStop}%
\bibitem [{\citenamefont {Lenosky}\ \emph {et~al.}(1992)\citenamefont
  {Lenosky}, \citenamefont {Gonze}, \citenamefont {Teter},\ and\ \citenamefont
  {Elser}}]{Lenosky}%
  \BibitemOpen
  \bibfield  {author} {\bibinfo {author} {\bibfnamefont {T.}~\bibnamefont
  {Lenosky}}, \bibinfo {author} {\bibfnamefont {X.}~\bibnamefont {Gonze}},
  \bibinfo {author} {\bibfnamefont {M.}~\bibnamefont {Teter}},\ and\ \bibinfo
  {author} {\bibfnamefont {V.}~\bibnamefont {Elser}},\ }\bibfield  {title}
  {\bibinfo {title} {Energetics of negatively curved graphitic carbon},\ }\href
  {https://doi.org/10.1038/355333a0} {\bibfield  {journal} {\bibinfo  {journal}
  {Nature}\ }\textbf {\bibinfo {volume} {355}},\ \bibinfo {pages} {333}
  (\bibinfo {year} {1992})}\BibitemShut {NoStop}%
\bibitem [{\citenamefont {Kas}\ \emph {et~al.}(2021)\citenamefont {Kas},
  \citenamefont {Vila}, \citenamefont {Rehr}, \citenamefont {Pemmaraju},\ and\
  \citenamefont {Tan}}]{FEFF10}%
  \BibitemOpen
  \bibfield  {author} {\bibinfo {author} {\bibfnamefont {J.~J.}\ \bibnamefont
  {Kas}}, \bibinfo {author} {\bibfnamefont {F.~D.}\ \bibnamefont {Vila}},
  \bibinfo {author} {\bibfnamefont {J.~J.}\ \bibnamefont {Rehr}}, \bibinfo
  {author} {\bibfnamefont {C.~D.}\ \bibnamefont {Pemmaraju}},\ and\ \bibinfo
  {author} {\bibfnamefont {T.~S.}\ \bibnamefont {Tan}},\ }\href
  {https://doi.org/10.48550/ARXIV.2106.13334} {\bibinfo {title} {Advanced
  calculations of x-ray spectroscopies with feff10 and corvus}} (\bibinfo
  {year} {2021})\BibitemShut {NoStop}%
\bibitem [{\citenamefont {Kaiser}(1966)}]{kaiser}%
  \BibitemOpen
  \bibfield  {author} {\bibinfo {author} {\bibfnamefont {J.~F.}\ \bibnamefont
  {Kaiser}},\ }\href@noop {} {\emph {\bibinfo {title} {Digital Filters” - Ch
  7 in Systems analysis by digital computer}}}\ (\bibinfo  {publisher} {John
  Wiley and Sons, New York},\ \bibinfo {year} {1966})\ pp.\ \bibinfo {pages}
  {218--285}\BibitemShut {NoStop}%
\bibitem [{\citenamefont {Comelli}\ \emph {et~al.}(1988)\citenamefont
  {Comelli}, \citenamefont {St\"ohr}, \citenamefont {Jark},\ and\ \citenamefont
  {Pate}}]{comelli}%
  \BibitemOpen
  \bibfield  {author} {\bibinfo {author} {\bibfnamefont {G.}~\bibnamefont
  {Comelli}}, \bibinfo {author} {\bibfnamefont {J.}~\bibnamefont {St\"ohr}},
  \bibinfo {author} {\bibfnamefont {W.}~\bibnamefont {Jark}},\ and\ \bibinfo
  {author} {\bibfnamefont {B.~B.}\ \bibnamefont {Pate}},\ }\bibfield  {title}
  {\bibinfo {title} {Extended x-ray-absorption fine-structure studies of
  diamond and graphite},\ }\href {https://doi.org/10.1103/PhysRevB.37.4383}
  {\bibfield  {journal} {\bibinfo  {journal} {Phys. Rev. B}\ }\textbf {\bibinfo
  {volume} {37}},\ \bibinfo {pages} {4383} (\bibinfo {year}
  {1988})}\BibitemShut {NoStop}%
\bibitem [{\citenamefont {Tanaka}\ \emph {et~al.}(2001)\citenamefont {Tanaka},
  \citenamefont {Matsubayashi}, \citenamefont {Imamura},\ and\ \citenamefont
  {Shimada}}]{Tanaka}%
  \BibitemOpen
  \bibfield  {author} {\bibinfo {author} {\bibfnamefont {T.}~\bibnamefont
  {Tanaka}}, \bibinfo {author} {\bibfnamefont {N.}~\bibnamefont
  {Matsubayashi}}, \bibinfo {author} {\bibfnamefont {M.}~\bibnamefont
  {Imamura}},\ and\ \bibinfo {author} {\bibfnamefont {H.}~\bibnamefont
  {Shimada}},\ }\bibfield  {title} {\bibinfo {title} {{Synchronous scanning of
  undulator gap and monochromator for XAFS measurement in soft X-ray region}},\
  }\href {https://doi.org/10.1107/S090904950001414X} {\bibfield  {journal}
  {\bibinfo  {journal} {Journal of Synchrotron Radiation}\ }\textbf {\bibinfo
  {volume} {8}},\ \bibinfo {pages} {345} (\bibinfo {year} {2001})}\BibitemShut
  {NoStop}%
\bibitem [{\citenamefont {Buades}\ \emph {et~al.}(2018)\citenamefont {Buades},
  \citenamefont {Moonshiram}, \citenamefont {Sidiropoulos}, \citenamefont
  {Le\'{o}n}, \citenamefont {Schmidt}, \citenamefont {Pi}, \citenamefont
  {Palo}, \citenamefont {Cousin}, \citenamefont {Pic\'{o}n}, \citenamefont
  {Koppens},\ and\ \citenamefont {Biegert}}]{Buades}%
  \BibitemOpen
  \bibfield  {author} {\bibinfo {author} {\bibfnamefont {B.}~\bibnamefont
  {Buades}}, \bibinfo {author} {\bibfnamefont {D.}~\bibnamefont {Moonshiram}},
  \bibinfo {author} {\bibfnamefont {T.~P.~H.}\ \bibnamefont {Sidiropoulos}},
  \bibinfo {author} {\bibfnamefont {I.}~\bibnamefont {Le\'{o}n}}, \bibinfo
  {author} {\bibfnamefont {P.}~\bibnamefont {Schmidt}}, \bibinfo {author}
  {\bibfnamefont {I.}~\bibnamefont {Pi}}, \bibinfo {author} {\bibfnamefont
  {N.~D.}\ \bibnamefont {Palo}}, \bibinfo {author} {\bibfnamefont {S.~L.}\
  \bibnamefont {Cousin}}, \bibinfo {author} {\bibfnamefont {A.}~\bibnamefont
  {Pic\'{o}n}}, \bibinfo {author} {\bibfnamefont {F.}~\bibnamefont {Koppens}},\
  and\ \bibinfo {author} {\bibfnamefont {J.}~\bibnamefont {Biegert}},\
  }\bibfield  {title} {\bibinfo {title} {Dispersive soft x-ray absorption
  fine-structure spectroscopy in graphite with an attosecond pulse},\ }\href
  {https://doi.org/10.1364/OPTICA.5.000502} {\bibfield  {journal} {\bibinfo
  {journal} {Optica}\ }\textbf {\bibinfo {volume} {5}},\ \bibinfo {pages} {502}
  (\bibinfo {year} {2018})}\BibitemShut {NoStop}%
\bibitem [{\citenamefont {Drabold}\ \emph {et~al.}(1995)\citenamefont
  {Drabold}, \citenamefont {Ordejón}, \citenamefont {Dong},\ and\
  \citenamefont {Martin}}]{DRABOLD1995833}%
  \BibitemOpen
  \bibfield  {author} {\bibinfo {author} {\bibfnamefont {D.~A.}\ \bibnamefont
  {Drabold}}, \bibinfo {author} {\bibfnamefont {P.}~\bibnamefont {Ordejón}},
  \bibinfo {author} {\bibfnamefont {J.}~\bibnamefont {Dong}},\ and\ \bibinfo
  {author} {\bibfnamefont {R.~M.}\ \bibnamefont {Martin}},\ }\bibfield  {title}
  {\bibinfo {title} {Spectral properties of large fullerenes: From cluster to
  crystal},\ }\href
  {https://doi.org/https://doi.org/10.1016/0038-1098(95)00562-5} {\bibfield
  {journal} {\bibinfo  {journal} {Solid State Communications}\ }\textbf
  {\bibinfo {volume} {96}},\ \bibinfo {pages} {833} (\bibinfo {year}
  {1995})}\BibitemShut {NoStop}%
\bibitem [{\citenamefont {Prasai}\ \emph {et~al.}(2018)\citenamefont {Prasai},
  \citenamefont {Subedi}, \citenamefont {Ferris}, \citenamefont {Biswas},\ and\
  \citenamefont {Drabold}}]{SPC}%
  \BibitemOpen
  \bibfield  {author} {\bibinfo {author} {\bibfnamefont {K.}~\bibnamefont
  {Prasai}}, \bibinfo {author} {\bibfnamefont {K.~N.}\ \bibnamefont {Subedi}},
  \bibinfo {author} {\bibfnamefont {K.}~\bibnamefont {Ferris}}, \bibinfo
  {author} {\bibfnamefont {P.}~\bibnamefont {Biswas}},\ and\ \bibinfo {author}
  {\bibfnamefont {D.~A.}\ \bibnamefont {Drabold}},\ }\bibfield  {title}
  {\bibinfo {title} {{Spatial Projection of Electronic Conductivity: The
  Example of Conducting Bridge Memory Materials}},\ }\href
  {https://doi.org/https://doi.org/10.1002/pssr.201800238} {\bibfield
  {journal} {\bibinfo  {journal} {physica status solidi (RRL) – Rapid
  Research Letters}\ }\textbf {\bibinfo {volume} {12}},\ \bibinfo {pages}
  {1800238} (\bibinfo {year} {2018})}\BibitemShut {NoStop}%
\bibitem [{\citenamefont {Allen}\ and\ \citenamefont
  {Feldman}(1993)}]{phonons1}%
  \BibitemOpen
  \bibfield  {author} {\bibinfo {author} {\bibfnamefont {P.~B.}\ \bibnamefont
  {Allen}}\ and\ \bibinfo {author} {\bibfnamefont {J.~L.}\ \bibnamefont
  {Feldman}},\ }\bibfield  {title} {\bibinfo {title} {Thermal conductivity of
  disordered harmonic solids},\ }\href
  {https://doi.org/10.1103/PhysRevB.48.12581} {\bibfield  {journal} {\bibinfo
  {journal} {Phys. Rev. B}\ }\textbf {\bibinfo {volume} {48}},\ \bibinfo
  {pages} {12581} (\bibinfo {year} {1993})}\BibitemShut {NoStop}%
\bibitem [{\citenamefont {Allen}\ \emph
  {et~al.}(1999{\natexlab{a}})\citenamefont {Allen}, \citenamefont {Feldman},
  \citenamefont {Fabian},\ and\ \citenamefont {Wooten}}]{phonon2}%
  \BibitemOpen
  \bibfield  {author} {\bibinfo {author} {\bibfnamefont {P.~B.}\ \bibnamefont
  {Allen}}, \bibinfo {author} {\bibfnamefont {J.~L.}\ \bibnamefont {Feldman}},
  \bibinfo {author} {\bibfnamefont {J.}~\bibnamefont {Fabian}},\ and\ \bibinfo
  {author} {\bibfnamefont {F.}~\bibnamefont {Wooten}},\ }\bibfield  {title}
  {\bibinfo {title} {Diffusons, locons and propagons: Character of atomie
  yibrations in amorphous si},\ }\href
  {https://doi.org/10.1080/13642819908223054} {\bibfield  {journal} {\bibinfo
  {journal} {Philosophical Magazine B}\ }\textbf {\bibinfo {volume} {79}},\
  \bibinfo {pages} {1715} (\bibinfo {year} {1999}{\natexlab{a}})}\BibitemShut
  {NoStop}%
\bibitem [{\citenamefont {Bell}\ and\ \citenamefont
  {Hibbins-Butler}(1975)}]{Bell_1975}%
  \BibitemOpen
  \bibfield  {author} {\bibinfo {author} {\bibfnamefont {R.~J.}\ \bibnamefont
  {Bell}}\ and\ \bibinfo {author} {\bibfnamefont {D.~C.}\ \bibnamefont
  {Hibbins-Butler}},\ }\bibfield  {title} {\bibinfo {title} {Acoustic and
  optical modes in vitreous silica, germania and beryllium fluoride},\ }\href
  {https://doi.org/10.1088/0022-3719/8/6/009} {\bibfield  {journal} {\bibinfo
  {journal} {Journal of Physics C: Solid State Physics}\ }\textbf {\bibinfo
  {volume} {8}},\ \bibinfo {pages} {787} (\bibinfo {year} {1975})}\BibitemShut
  {NoStop}%
\bibitem [{\citenamefont {Allen}\ \emph
  {et~al.}(1999{\natexlab{b}})\citenamefont {Allen}, \citenamefont {Feldman},
  \citenamefont {Fabian},\ and\ \citenamefont {Wooten}}]{Allen_Feldman}%
  \BibitemOpen
  \bibfield  {author} {\bibinfo {author} {\bibfnamefont {P.~B.}\ \bibnamefont
  {Allen}}, \bibinfo {author} {\bibfnamefont {J.~L.}\ \bibnamefont {Feldman}},
  \bibinfo {author} {\bibfnamefont {J.}~\bibnamefont {Fabian}},\ and\ \bibinfo
  {author} {\bibfnamefont {F.}~\bibnamefont {Wooten}},\ }\bibfield  {title}
  {\bibinfo {title} {Diffusons, locons and propagons: Character of atomie
  yibrations in amorphous si},\ }\href
  {https://doi.org/10.1080/13642819908223054} {\bibfield  {journal} {\bibinfo
  {journal} {Philosophical Magazine B}\ }\textbf {\bibinfo {volume} {79}},\
  \bibinfo {pages} {1715} (\bibinfo {year} {1999}{\natexlab{b}})},\ \Eprint
  {https://arxiv.org/abs/https://doi.org/10.1080/13642819908223054}
  {https://doi.org/10.1080/13642819908223054} \BibitemShut {NoStop}%
\bibitem [{\citenamefont {Seyf}\ \emph {et~al.}(2017)\citenamefont {Seyf},
  \citenamefont {Yates}, \citenamefont {Bougher}, \citenamefont {Graham},
  \citenamefont {Baratunde A.~Cola}, \citenamefont {Ji}, \citenamefont {Kim},
  \citenamefont {Dupuis}, \citenamefont {Lv},\ and\ \citenamefont
  {Henry}}]{Henry35}%
  \BibitemOpen
  \bibfield  {author} {\bibinfo {author} {\bibfnamefont {H.~R.}\ \bibnamefont
  {Seyf}}, \bibinfo {author} {\bibfnamefont {L.}~\bibnamefont {Yates}},
  \bibinfo {author} {\bibfnamefont {T.~L.}\ \bibnamefont {Bougher}}, \bibinfo
  {author} {\bibfnamefont {S.}~\bibnamefont {Graham}}, \bibinfo {author}
  {\bibfnamefont {T.~D.}\ \bibnamefont {Baratunde A.~Cola}}, \bibinfo {author}
  {\bibfnamefont {M.-H.}\ \bibnamefont {Ji}}, \bibinfo {author} {\bibfnamefont
  {J.}~\bibnamefont {Kim}}, \bibinfo {author} {\bibfnamefont {R.}~\bibnamefont
  {Dupuis}}, \bibinfo {author} {\bibfnamefont {W.}~\bibnamefont {Lv}},\ and\
  \bibinfo {author} {\bibfnamefont {A.}~\bibnamefont {Henry}},\ }\bibfield
  {title} {\bibinfo {title} {Rethinking phonons: The issue of disorder},\
  }\href {https://doi.org/10.1038/s41524-017-0052-9} {\bibfield  {journal}
  {\bibinfo  {journal} {npj Computational Materials}\ }\textbf {\bibinfo
  {volume} {3}},\ \bibinfo {pages} {49} (\bibinfo {year} {2017})}\BibitemShut
  {NoStop}%
\bibitem [{\citenamefont {Feldman}\ \emph {et~al.}(1999)\citenamefont
  {Feldman}, \citenamefont {Allen},\ and\ \citenamefont {Bickham}}]{Henry36}%
  \BibitemOpen
  \bibfield  {author} {\bibinfo {author} {\bibfnamefont {J.~L.}\ \bibnamefont
  {Feldman}}, \bibinfo {author} {\bibfnamefont {P.~B.}\ \bibnamefont {Allen}},\
  and\ \bibinfo {author} {\bibfnamefont {S.~R.}\ \bibnamefont {Bickham}},\
  }\bibfield  {title} {\bibinfo {title} {Numerical study of low-frequency
  vibrations in amorphous silicon},\ }\href
  {https://doi.org/10.1103/PhysRevB.59.3551} {\bibfield  {journal} {\bibinfo
  {journal} {Phys. Rev. B}\ }\textbf {\bibinfo {volume} {59}},\ \bibinfo
  {pages} {3551} (\bibinfo {year} {1999})}\BibitemShut {NoStop}%
\bibitem [{\citenamefont {Gurney}(1952)}]{vib1}%
  \BibitemOpen
  \bibfield  {author} {\bibinfo {author} {\bibfnamefont {R.~W.}\ \bibnamefont
  {Gurney}},\ }\bibfield  {title} {\bibinfo {title} {Lattice vibrations in
  graphite},\ }\href {https://doi.org/10.1103/PhysRev.88.465} {\bibfield
  {journal} {\bibinfo  {journal} {Phys. Rev.}\ }\textbf {\bibinfo {volume}
  {88}},\ \bibinfo {pages} {465} (\bibinfo {year} {1952})}\BibitemShut
  {NoStop}%
\bibitem [{\citenamefont {Sherry}\ and\ \citenamefont {Coulson}(1956)}]{vib2}%
  \BibitemOpen
  \bibfield  {author} {\bibinfo {author} {\bibfnamefont {P.~B.}\ \bibnamefont
  {Sherry}}\ and\ \bibinfo {author} {\bibfnamefont {C.~A.}\ \bibnamefont
  {Coulson}},\ }\bibfield  {title} {\bibinfo {title} {The vibrational frequency
  distribution of graphite: I. out-of-plane modes of a single layer},\ }\href
  {https://doi.org/10.1088/0370-1301/69/12/319} {\bibfield  {journal} {\bibinfo
   {journal} {Proceedings of the Physical Society. Section B}\ }\textbf
  {\bibinfo {volume} {69}},\ \bibinfo {pages} {1326} (\bibinfo {year}
  {1956})}\BibitemShut {NoStop}%
\bibitem [{\citenamefont {Newell}(1957)}]{vib3}%
  \BibitemOpen
  \bibfield  {author} {\bibinfo {author} {\bibfnamefont {G.~F.}\ \bibnamefont
  {Newell}},\ }\bibfield  {title} {\bibinfo {title} {Vibration spectrum of
  graphite and boron nitride. ii. the three‐dimensional spectrum},\ }\href
  {https://doi.org/10.1063/1.1743680} {\bibfield  {journal} {\bibinfo
  {journal} {The Journal of Chemical Physics}\ }\textbf {\bibinfo {volume}
  {27}},\ \bibinfo {pages} {240} (\bibinfo {year} {1957})},\ \Eprint
  {https://arxiv.org/abs/https://doi.org/10.1063/1.1743680}
  {https://doi.org/10.1063/1.1743680} \BibitemShut {NoStop}%
\bibitem [{\citenamefont {Lui}\ \emph {et~al.}(2012)\citenamefont {Lui},
  \citenamefont {Malard}, \citenamefont {Kim}, \citenamefont {Lantz},
  \citenamefont {Laverge}, \citenamefont {Saito},\ and\ \citenamefont
  {Heinz}}]{vib4}%
  \BibitemOpen
  \bibfield  {author} {\bibinfo {author} {\bibfnamefont {C.~H.}\ \bibnamefont
  {Lui}}, \bibinfo {author} {\bibfnamefont {L.~M.}\ \bibnamefont {Malard}},
  \bibinfo {author} {\bibfnamefont {S.}~\bibnamefont {Kim}}, \bibinfo {author}
  {\bibfnamefont {G.}~\bibnamefont {Lantz}}, \bibinfo {author} {\bibfnamefont
  {F.~E.}\ \bibnamefont {Laverge}}, \bibinfo {author} {\bibfnamefont
  {R.}~\bibnamefont {Saito}},\ and\ \bibinfo {author} {\bibfnamefont {T.~F.}\
  \bibnamefont {Heinz}},\ }\bibfield  {title} {\bibinfo {title} {Observation of
  layer-breathing mode vibrations in few-layer graphene through combination
  raman scattering},\ }\href {https://doi.org/10.1021/nl302450s} {\bibfield
  {journal} {\bibinfo  {journal} {Nano Letters}\ }\textbf {\bibinfo {volume}
  {12}},\ \bibinfo {pages} {5539} (\bibinfo {year} {2012})},\ \bibinfo {note}
  {pMID: 22963681},\ \Eprint
  {https://arxiv.org/abs/https://doi.org/10.1021/nl302450s}
  {https://doi.org/10.1021/nl302450s} \BibitemShut {NoStop}%
\bibitem [{\citenamefont {Seyf}\ \emph {et~al.}(2018)\citenamefont {Seyf},
  \citenamefont {Lv}, \citenamefont {Rohskopf},\ and\ \citenamefont
  {Henry}}]{Henry}%
  \BibitemOpen
  \bibfield  {author} {\bibinfo {author} {\bibfnamefont {H.~R.}\ \bibnamefont
  {Seyf}}, \bibinfo {author} {\bibfnamefont {W.}~\bibnamefont {Lv}}, \bibinfo
  {author} {\bibfnamefont {A.}~\bibnamefont {Rohskopf}},\ and\ \bibinfo
  {author} {\bibfnamefont {A.}~\bibnamefont {Henry}},\ }\bibfield  {title}
  {\bibinfo {title} {The importance of phonons with negative phase quotient in
  disordered solids},\ }\href {https://doi.org/10.1038/s41598-018-20704-7}
  {\bibfield  {journal} {\bibinfo  {journal} {Scientific reports}\ }\textbf
  {\bibinfo {volume} {8}},\ \bibinfo {pages} {2627} (\bibinfo {year}
  {2018})}\BibitemShut {NoStop}%
\bibitem [{\citenamefont {Surblys}\ \emph {et~al.}(2021)\citenamefont
  {Surblys}, \citenamefont {Matsubara}, \citenamefont {Kikugawa},\ and\
  \citenamefont {Ohara}}]{heatflux1}%
  \BibitemOpen
  \bibfield  {author} {\bibinfo {author} {\bibfnamefont {D.}~\bibnamefont
  {Surblys}}, \bibinfo {author} {\bibfnamefont {H.}~\bibnamefont {Matsubara}},
  \bibinfo {author} {\bibfnamefont {G.}~\bibnamefont {Kikugawa}},\ and\
  \bibinfo {author} {\bibfnamefont {T.}~\bibnamefont {Ohara}},\ }\bibfield
  {title} {\bibinfo {title} {Methodology and meaning of computing heat flux via
  atomic stress in systems with constraint dynamics},\ }\href
  {https://doi.org/10.1063/5.0070930} {\bibfield  {journal} {\bibinfo
  {journal} {Journal of Applied Physics}\ }\textbf {\bibinfo {volume} {130}},\
  \bibinfo {pages} {215104} (\bibinfo {year} {2021})},\ \Eprint
  {https://arxiv.org/abs/https://doi.org/10.1063/5.0070930}
  {https://doi.org/10.1063/5.0070930} \BibitemShut {NoStop}%
\bibitem [{\citenamefont {Surblys}\ \emph {et~al.}(2019)\citenamefont
  {Surblys}, \citenamefont {Matsubara}, \citenamefont {Kikugawa},\ and\
  \citenamefont {Ohara}}]{heatflux2}%
  \BibitemOpen
  \bibfield  {author} {\bibinfo {author} {\bibfnamefont {D.}~\bibnamefont
  {Surblys}}, \bibinfo {author} {\bibfnamefont {H.}~\bibnamefont {Matsubara}},
  \bibinfo {author} {\bibfnamefont {G.}~\bibnamefont {Kikugawa}},\ and\
  \bibinfo {author} {\bibfnamefont {T.}~\bibnamefont {Ohara}},\ }\bibfield
  {title} {\bibinfo {title} {Application of atomic stress to compute heat flux
  via molecular dynamics for systems with many-body interactions},\ }\href
  {https://doi.org/10.1103/PhysRevE.99.051301} {\bibfield  {journal} {\bibinfo
  {journal} {Phys. Rev. E}\ }\textbf {\bibinfo {volume} {99}},\ \bibinfo
  {pages} {051301} (\bibinfo {year} {2019})}\BibitemShut {NoStop}%
\bibitem [{\citenamefont {Green}(1954)}]{Green}%
  \BibitemOpen
  \bibfield  {author} {\bibinfo {author} {\bibfnamefont {M.~S.}\ \bibnamefont
  {Green}},\ }\bibfield  {title} {\bibinfo {title} {Markoff random processes
  and the statistical mechanics of time‐dependent phenomena. ii. irreversible
  processes in fluids},\ }\href {https://doi.org/10.1063/1.1740082} {\bibfield
  {journal} {\bibinfo  {journal} {The Journal of Chemical Physics}\ }\textbf
  {\bibinfo {volume} {22}},\ \bibinfo {pages} {398} (\bibinfo {year} {1954})},\
  \Eprint {https://arxiv.org/abs/https://doi.org/10.1063/1.1740082}
  {https://doi.org/10.1063/1.1740082} \BibitemShut {NoStop}%
\bibitem [{\citenamefont {Kubo}\ \emph {et~al.}(1957)\citenamefont {Kubo},
  \citenamefont {Yokota},\ and\ \citenamefont {Nakajima}}]{Kubo}%
  \BibitemOpen
  \bibfield  {author} {\bibinfo {author} {\bibfnamefont {R.}~\bibnamefont
  {Kubo}}, \bibinfo {author} {\bibfnamefont {M.}~\bibnamefont {Yokota}},\ and\
  \bibinfo {author} {\bibfnamefont {S.}~\bibnamefont {Nakajima}},\ }\bibfield
  {title} {\bibinfo {title} {Statistical-mechanical theory of irreversible
  processes. ii. response to thermal disturbance},\ }\href
  {https://doi.org/10.1143/JPSJ.12.1203} {\bibfield  {journal} {\bibinfo
  {journal} {Journal of the Physical Society of Japan}\ }\textbf {\bibinfo
  {volume} {12}},\ \bibinfo {pages} {1203} (\bibinfo {year} {1957})},\ \Eprint
  {https://arxiv.org/abs/https://doi.org/10.1143/JPSJ.12.1203}
  {https://doi.org/10.1143/JPSJ.12.1203} \BibitemShut {NoStop}%
\bibitem [{\citenamefont {Nosé}(1984)}]{nose}%
  \BibitemOpen
  \bibfield  {author} {\bibinfo {author} {\bibfnamefont {S.}~\bibnamefont
  {Nosé}},\ }\bibfield  {title} {\bibinfo {title} {A molecular dynamics method
  for simulations in the canonical ensemble},\ }\href
  {https://doi.org/10.1080/00268978400101201} {\bibfield  {journal} {\bibinfo
  {journal} {Molecular Physics}\ }\textbf {\bibinfo {volume} {52}},\ \bibinfo
  {pages} {255} (\bibinfo {year} {1984})},\ \Eprint
  {https://arxiv.org/abs/https://doi.org/10.1080/00268978400101201}
  {https://doi.org/10.1080/00268978400101201} \BibitemShut {NoStop}%
\bibitem [{\citenamefont {Hoover}(1985)}]{hoover}%
  \BibitemOpen
  \bibfield  {author} {\bibinfo {author} {\bibfnamefont {W.~G.}\ \bibnamefont
  {Hoover}},\ }\bibfield  {title} {\bibinfo {title} {{Canonical dynamics:
  Equilibrium phase-space distributions}},\ }\href
  {https://doi.org/10.1103/PhysRevA.31.1695} {\bibfield  {journal} {\bibinfo
  {journal} {Phys. Rev. A}\ }\textbf {\bibinfo {volume} {31}},\ \bibinfo
  {pages} {1695} (\bibinfo {year} {1985})}\BibitemShut {NoStop}%
\bibitem [{\citenamefont {Allen}\ and\ \citenamefont
  {Feldman}(1989)}]{Allen_Feldman_aSi}%
  \BibitemOpen
  \bibfield  {author} {\bibinfo {author} {\bibfnamefont {P.~B.}\ \bibnamefont
  {Allen}}\ and\ \bibinfo {author} {\bibfnamefont {J.~L.}\ \bibnamefont
  {Feldman}},\ }\bibfield  {title} {\bibinfo {title} {Thermal conductivity of
  glasses: Theory and application to amorphous si},\ }\href
  {https://doi.org/10.1103/PhysRevLett.62.645} {\bibfield  {journal} {\bibinfo
  {journal} {Phys. Rev. Lett.}\ }\textbf {\bibinfo {volume} {62}},\ \bibinfo
  {pages} {645} (\bibinfo {year} {1989})}\BibitemShut {NoStop}%
\bibitem [{\citenamefont {Ho}\ \emph {et~al.}(1972)\citenamefont {Ho},
  \citenamefont {Powell},\ and\ \citenamefont {Liley}}]{TCofElements}%
  \BibitemOpen
  \bibfield  {author} {\bibinfo {author} {\bibfnamefont {C.~Y.}\ \bibnamefont
  {Ho}}, \bibinfo {author} {\bibfnamefont {R.~W.}\ \bibnamefont {Powell}},\
  and\ \bibinfo {author} {\bibfnamefont {P.~E.}\ \bibnamefont {Liley}},\
  }\bibfield  {title} {\bibinfo {title} {Thermal conductivity of the
  elements},\ }\href {https://doi.org/10.1063/1.3253100} {\bibfield  {journal}
  {\bibinfo  {journal} {Journal of Physical and Chemical Reference Data}\
  }\textbf {\bibinfo {volume} {1}},\ \bibinfo {pages} {279} (\bibinfo {year}
  {1972})},\ \Eprint {https://arxiv.org/abs/https://doi.org/10.1063/1.3253100}
  {https://doi.org/10.1063/1.3253100} \BibitemShut {NoStop}%
\bibitem [{\citenamefont {Stukowski}(2009)}]{OVITO}%
  \BibitemOpen
  \bibfield  {author} {\bibinfo {author} {\bibfnamefont {A.}~\bibnamefont
  {Stukowski}},\ }\bibfield  {title} {\bibinfo {title} {Visualization and
  analysis of atomistic simulation data with {OVITO}{\textendash}the open
  visualization tool},\ }\href {https://doi.org/10.1088/0965-0393/18/1/015012}
  {\bibfield  {journal} {\bibinfo  {journal} {Modelling and Simulation in
  Materials Science and Engineering}\ }\textbf {\bibinfo {volume} {18}},\
  \bibinfo {pages} {015012} (\bibinfo {year} {2009})}\BibitemShut {NoStop}%
\bibitem [{Jmo()}]{Jmol}%
  \BibitemOpen
  \href {http://www.jmol.org/} {\bibinfo {title} {{Jmol: an open-source Java
  viewer for chemical structures in 3D.}}}\BibitemShut {Stop}%
\end{thebibliography}%


\begin{thebibliography}{0}%
\makeatletter
\providecommand \@ifxundefined [1]{%
 \@ifx{#1\undefined}
}%
\providecommand \@ifnum [1]{%
 \ifnum #1\expandafter \@firstoftwo
 \else \expandafter \@secondoftwo
 \fi
}%
\providecommand \@ifx [1]{%
 \ifx #1\expandafter \@firstoftwo
 \else \expandafter \@secondoftwo
 \fi
}%
\providecommand \natexlab [1]{#1}%
\providecommand \enquote  [1]{``#1''}%
\providecommand \bibnamefont  [1]{#1}%
\providecommand \bibfnamefont [1]{#1}%
\providecommand \citenamefont [1]{#1}%
\providecommand \href@noop [0]{\@secondoftwo}%
\providecommand \href [0]{\begingroup \@sanitize@url \@href}%
\providecommand \@href[1]{\@@startlink{#1}\@@href}%
\providecommand \@@href[1]{\endgroup#1\@@endlink}%
\providecommand \@sanitize@url [0]{\catcode `\\12\catcode `\$12\catcode
  `\&12\catcode `\#12\catcode `\^12\catcode `\_12\catcode `\%12\relax}%
\providecommand \@@startlink[1]{}%
\providecommand \@@endlink[0]{}%
\providecommand \url  [0]{\begingroup\@sanitize@url \@url }%
\providecommand \@url [1]{\endgroup\@href {#1}{\urlprefix }}%
\providecommand \urlprefix  [0]{URL }%
\providecommand \Eprint [0]{\href }%
\providecommand \doibase [0]{https://doi.org/}%
\providecommand \selectlanguage [0]{\@gobble}%
\providecommand \bibinfo  [0]{\@secondoftwo}%
\providecommand \bibfield  [0]{\@secondoftwo}%
\providecommand \translation [1]{[#1]}%
\providecommand \BibitemOpen [0]{}%
\providecommand \bibitemStop [0]{}%
\providecommand \bibitemNoStop [0]{.\EOS\space}%
\providecommand \EOS [0]{\spacefactor3000\relax}%
\providecommand \BibitemShut  [1]{\csname bibitem#1\endcsname}%
\let\auto@bib@innerbib\@empty
\end{thebibliography}%

\end{document}



\title{Supplementary Material: Atomistic Nature of Amorphous Graphite}

\author{C. Ugwumadu}
\email{Corresponding author.\\ E-mail: cu884120@ohio.edu}
\affiliation{Department of Physics and Astronomy, \\
Nanoscale and Quantum Phenomena Institute (NQPI),\\
Ohio University, Athens, Ohio 45701, USA}%

\author{K. Nepal}
\affiliation{Department of Physics and Astronomy, \\
Nanoscale and Quantum Phenomena Institute (NQPI),\\
Ohio University, Athens, Ohio 45701, USA}%

\author{R. Thapa}
\affiliation{Department of Physics and Astronomy, \\
Nanoscale and Quantum Phenomena Institute (NQPI),\\
Ohio University, Athens, Ohio 45701, USA}%

\author{D. A. Drabold}
\affiliation{Department of Physics and Astronomy, \\
Nanoscale and Quantum Phenomena Institute (NQPI),\\
Ohio University, Athens, Ohio 45701, USA}%

\date{\today}

\maketitle

\section{Description of Animations produced for the amorphous Graphen (aG) models}
\begin{figure*}[!htpb]
	\centering
	\includegraphics[width=.5\textwidth]{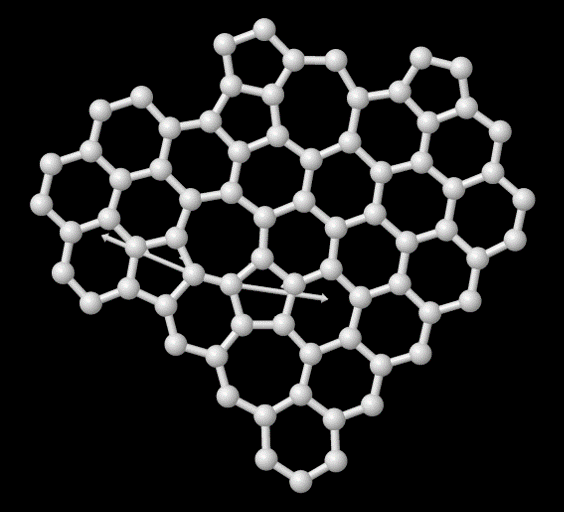}
    	\caption{Snapshot of a high-frequency vibration in aG. The arrows show the direction of motion for the participating atoms.}
	\label{fig:CSfig_OpticMode}
\end{figure*}

We have produced some animations to aid the reader in visualizing some of the discussions in the paper. The animations can be found \href{https://people.ohio.edu/drabold/kent_movies/}{here} or by visiting the url: \url{https://people.ohio.edu/drabold/kent_movies/}

The descriptions for the video files are as follows:

\begin{enumerate}
  \item \textbf{aG\_Formation.mp4}: This describes the formation process of aG. 
  \item \textbf{high\_freq.mp4}: This shows the vibration in aG at the high-frequency end of the spectrum, where the localization is on pentagon ring-forming atoms. Fig. \ref{fig:CSfig_OpticMode} is a snapshot of one is such vibrations, the arrows in Fig. \ref{fig:CSfig_OpticMode} (and subsequent animations) is a vector indicating the vibration path for the participating atoms. The animation show the vibration at a frequency of 2023.13 cm${-1}$
  \item \textbf{quasiLocalized\_freq.mp4}  : This vibration has a mid-spectrum frequency of 745.32 cm$^{-1}$. It shows the resonant modes in the vibrations of aG. We draw the reader's attention to the mixture of in-plane and out-of-plane vibrations at this frequency.
  \item \textbf{low\_freq.mp4}: The animation show vibrations at the low-end frequency of 76.54 cm${-1}$ is shown in this animation. clearly, the vibrations are not localized and constitute mostly out-of-plane, "acoustic-like" vibrations

\end{enumerate}

